\documentstyle[aps,twocolumn,epsf]{revtex} %%\tightenlines

\def\lsim{\lower.8ex\hbox{$\buildrel<\over\sim$}}
\def\gsim{\lower.8ex\hbox{$\buildrel>\over\sim$}}
\begin{document} 
\draft
\twocolumn[\hsize\textwidth\columnwidth\hsize\csname
@twocolumnfalse\endcsname
%\tightenlines
\title{ Influence of solvent granularity on the effective 
interaction between charged colloidal suspensions}
\author{E. Allahyarov  and H. L{\"o}wen}

\address{Institut f{\"u}r Theoretische Physik II, 
Heinrich-Heine-Universit{\"a}t D{\"u}sseldorf, D-40225 D{\"u}sseldorf,
Germany}
\date{\today}

\maketitle
\begin{abstract}

 We study the effect of solvent granularity on the
effective force  between two charged colloidal particles by computer
simulations of the primitive model of strongly asymmetric electrolytes
with an explicitly added hard sphere
solvent. Apart from molecular oscillating forces for nearly
touching colloids which arise from solvent
and counterion layering, the counterions are attracted towards the colloidal surfaces
by solvent depletion  providing a simple statistical description of hydration.
This, in turn, has an important influence  on the effective forces for
larger distances which are considerably reduced as compared to the prediction 
based on the primitive model. When these forces are repulsive,  the long-distance behaviour can be
described by an effective Yukawa pair potential with a solvent-renormalized charge.
As a function of colloidal volume fraction and added salt concentration, this
solvent-renormalized charge behaves qualitatively similar to that obtained
via the Poisson-Boltzmann cell model but there are quantitative differences. 
For divalent counterions and nano-sized colloids,
on the other hand, the hydration may lead to  overscreened colloids with mutual
attraction while the primitive model yields repulsive forces. All these
new effects can be accounted
for through  a solvent-averaged primitive model (SPM) which is obtained from the full
model by integrating out the solvent degrees of freedom. The SPM was  used
to access larger colloidal particles without simulating the solvent explicitly.
\end{abstract}
]
\renewcommand{\thepage}{\hskip 8.9cm \arabic{page} \hfill Typeset
using REV\TeX }
\narrowtext

\section{Introduction}
Most of soft matter systems, as colloids, polymers  or biological macromolecules,
are dispersed in a molecular solvent \cite{ref1}. Therefore, 
a full statistical description of supramolecular solutions
should include the solvent explicitly. 
Such a treatment is highly non-trivial, however, since the
length scale separation between the mesoscopic particles and the molecular solvent
directly implies that the number of solvent particles which have to be included
is enormous. On the other hand, one is interested mainly in properties
of the big particles such that a solvent pre-average makes sense. The
crudest form of such a course-grained level is to treat  solvent properties
just by a dielectric background or by some effective parameters which enter
in the effective colloidal interactions.  This procedure is questionable
for polyelectrolytes where the solvent couples directly to the counterions 
which may affect the effective interaction between the polyelectrolytes
via the long-ranged Coulomb coupling of counterions to the  polyelectrolytes.

In this paper, we consider the case of two spherical charged colloidal particles
(polyions) which are immersed in a bath of the molecular  solvent and
their oppositely charged counterions plus additional salt ions \cite{HansenLoewen}.
 Our main
focus is the total effective force acting onto the colloidal pair which is the key quantity
to understand colloidal stability and which governs colloidal correlations and phase
transitions. In almost any
theoretical treatment, the discrete structure of the solvent particles was neglected
and only the charged constituents  were treated explicitly within the so-called
``primitive" model (PM) of strongly asymmetric electrolytes. Even this model
is non-trivial in the colloidal context due to the large asymmetry between poly- and counterions
and bears a rich physics resulting from the strong coupling between the different species.
In recent computer simulations \cite{Hribar,Pincus,AllahyarovPRL,Linse,Messina}, counterionic 
correlations have been shown  to be responsible for effective attractions
between the like-charge polyions. The PM, re-formulated in terms of modern density functional 
theory of the inhomogeneous counterion plasma \cite{LHM}, can also  also used as a 
starting point to
derive simpler theories such as the mean-field nonlinear Poisson-Boltzmann 
approach or the linearized Debye-H\"uckel-type screening theory. The latter results
in an effective Yukawa pair potential between the colloids as
given by the electrostatic part of the celebrated
Derjaguin-Landau-Verwey-Overbeek (DLVO) theory \cite{Verwey}. This potential
can also be used with renormalized parameters to include parts of the nonlinear screening
effects arising from Poisson-Boltzmann theory \cite{Alexander}.

In the present paper, we investigate the influence
 of {\it solvent granularity\/}  on the effective interactions
between charged colloids. We model the solvent as a hard sphere
fluid at intermediate packing fractions and use computer simulations 
and the theoretical concept of
effective interactions to derive effects due to the discrete solvent.
The PM is tested against this more general model.
Although the hard sphere model  neglects some important solvent properties as 
its polarizibility \cite{Smith} and its permanent multipole  moments \cite{LIE2},
it provides a minimal framework to get insight into counterion hydration 
and screening effects. The hard sphere solvent 
 model (which is  sometimes called solvent-primitive model) has been used also in many other
investigations of ordinary electrolytes and for electrolytes confined between two
parallel charged plates. Most of the approaches invoke additional approximations
as different versions of liquid-integral equations \cite{LIE,LIE2,planar},
Poisson-Boltzmann theory suitably modified to include the short-ranged solvent depletion
effects \cite{MPB},
or more sophisticated density functional approaches of multicomponent systems \cite{DFT}. 
For charged plates \cite{Henderson2} and for small 
neutral particles \cite{CS} some computer simulations have already been
performed including  a hard sphere solvent explicitly 
but there are no results for charged colloidal spheres. 

Most of the results in this paper are based on 
 a new ``solvent bath" simulation scheme which allows to simulate many neutral spheres
together with the charged species.  For divalent counterions, we
 obtain attractive forces due to overscreening of
polyions by counterions which are attracted towards the colloidal surfaces via hydration 
(or solvent depletion) forces. For monovalent counterions and large distances we
show that the concept of charge renormalization can be used to extract a Yukawa picture
of the effective interaction with a solvent-renormalized polyion charge. 
We check the trends of this renormalized charge with respect to the colloidal density
and the concentration of added salt and find qualitative agreement but quantitative differences
as compared to the Poisson-Boltzmann theory.
All our results can be reproduced within a solvent-averaged 
primitive model (SPM) which was extensively used in earlier theoretical studies of electrolytes
between plates \cite{LIE,planar2}. This idea originates from McMillan
and Mayer \cite{MM}  dating back to 1945.

Our paper is organized as follows: In chapter II, we describe our model and define 
approximations on different levels.
The computer simulation method is described in chapter III. We then turn to
results for the neutral case in chapter IV and for the salt-free case in 
chapter V. Parts of the latter have been published elsewhere
\cite{El1}. The effect of added salt is described
in chapter VI and other mechanisms of polyion-polyion attraction
are critically discussed in chapter VII. We finally conclude in chapter VIII.

\section{Modeling on different levels}

In this section we summarize the modeling on different levels. In the following
we shall use the most detailed description of the hard sphere solvent model and test
the validity of the different inferior levels with respect to our data.

\subsection{The  hard sphere solvent model (HSSM)}

The hard sphere solvent model (HSSM) involves
 spherical polyions  with diameter $\sigma_p$ and homogeneously
smeared charge $q_p$ together with their  counterions of diameter $\sigma_c$ and charge $q_c$
in a bath of a neutral solvent $(q_s=0)$ with diameter $\sigma_s$. 
In the absence of salt, the pair potentials
between the particles as a function of their  mutual distances $r$ 
are  a combination of excluded volume and Coulomb terms
\begin{equation}
V_{ij} (r) = \cases { \infty &for $ r \leq (\sigma_i + \sigma_j)/2 $\cr
   {{q_iq_j} / {\epsilon r}} &else\cr}
\label{1}
\end{equation}
where $\epsilon$ is the a smeared background dielectric constant of the solvent
and $(ij)=(pp),(pc),(ps),(cc),(cs),(ss)$.
Further parameters are the thermal energy $k_BT$ and the partial number densities
$\rho_i \ \ (i=p,c,s)$ which can be expressed as partial volume fractions
$\phi_i= \pi \rho_i \sigma_i^3/6 \ (i=p,c,s)$.
Charge neutrality requires $\rho_p | q_p |= \rho_c | q_c |$. Additional
salt ions can readily be included into the description as further charged hard spheres.

\subsection{The solvent-averaged primitive model (SPM)}

For a fixed configuration of charged particles the solvent can be traced out exactly arriving
at depletion forces ${\vec F}^{(d)}_i$ 
acting onto the $i$th  charged particle. They can be related to
a surface integral over the
 solvent equilibrium density field $\rho_s({\vec r} )$ which depends parametrically
on the positions of the fixed charged particles:
\begin{equation}
{\vec F}^{(d)}_i=k_BT \int_{{\cal S}_i} d{\vec f} \ \ \rho_s({\vec r}) 
\label{9}
\end{equation}
where ${\vec f}$ is a surface  vector pointing towards the  center of the $i$th charged particle.
If one adds these forces to the PM, the resulting model is strictly equivalent to the HSSM.
The integrand $\rho_s({\vec r} )$ is affected by the space excluded for the
solvent due to the presence of the finite core of the charged particles resulting
in inhomogeneous density distributions around the excluded volume. The range
of this
inhomogeneity is characterized by the hard sphere bulk correlation length $\xi$
which depends on the solvent packing fraction $\phi_s$. 
A further approximation decomposes the forces ${\vec F}^{(d)}_i$ into pairwise parts, i.e.
into a superposition of pair contributions coming from neighboring charged particles.
This approximation is justified if the average distance between triplets, quadruplets, etc.\ of
charged particles is much larger than the bulk correlation length $\xi$. In the salt-free
case, this is generally granted except for nearly touching polyions with squeezed counterions.
If salt is added, the justification is less clear as ion pairing by counter- and coions 
near polyions may be an important configuration.

The resulting pairwise interactions  define the
solvent-averaged model (SPM) where the depletion pair potentials
$V^{(d)}_{ij}(r)$  ($(ij)=(pp),(pc),(cc)$) have to be added to the interactions of the primitive model
of the next paragraph. These pairwise  
 depletion forces have  been the subject of intense
recent research \cite{depl10,depl1a,depl1b,depl1c,depl2a,depl2b}. In particular,
we will determine them by computer simulation, and use these results as an input for the SPM.

\subsection{The primitive model (PM)}

The primitive model has the same interactions as the HSSM except for the absence of the solvent.
Hence the basic interactions are again
\begin{equation}
V_{ij} (r) = \cases { \infty &for $ r \leq (\sigma_i + \sigma_j)/2 $\cr
   {{q_iq_j} / {\epsilon r}} &else\cr}
\label{111}
\end{equation}
but now for $(ij)=(pp),(pc),(cc)$ only.

\subsection{DLVO-theory}

In DLVO theory, only the polyions are treated explicitly. The electrostatic part of 
their interaction is an effective Yukawa pair potential which 
has the form:
\begin{equation}
V(r)= {{q_p^2\exp ( -\kappa (r-\sigma_p ))}\over{
(1+\kappa\sigma_p/2)^2 \epsilon r }}
\label{2}
\end{equation}
with 

\begin{equation}
\kappa=\sqrt{4\pi\rho_c q_c^2/\epsilon k_BT}
\label{2a}
\end{equation}

\subsection{The PB-renormalized Yukawa model (PBYM)}

This approach was suggested by Alexander et al \cite{Alexander} and is based
on Poisson-Boltzmann theory in a spherical cell around a single polyion.
The cell  radius $R$ is fixed by the polyion concentration:
\begin{equation}
R=(4\pi\rho_p /3)^{-1/3}
\label{2b}
\end{equation}
Within Poisson-Boltzmann theory, one  calculates the counterion 
density ${\tilde \rho}_c$ at the cell boundary. Linearizing the nonlinear
Poisson-Boltzmann theory at the cell boundary,
one obtains again an effective Yukawa potential 
between the colloids arriving at the PB-renormalized Yukawa model (PBYM).
The Yukawa potential
has the same form as in Eqn.(\ref{2}) but contains a 
renormalized inverse screening length 
\begin{equation}
\kappa^* = \kappa \sqrt{{{\tilde \rho}_c} \over{\rho_c} } 
\label{2c}
\end{equation}
and  a {\it renormalized  charge}
\begin{equation}
q_p^*=q_p {{{\tilde \rho}_c} \over{\rho_c} }
\label{2d}
\end{equation}
which is considerably smaller than the bare charge $q_p$. 
Many experimental data
for the colloidal structural correlations \cite{Haertl}, the long-time
self-diffusion \cite{Palberg2}
or the freezing
line \cite{Palberg1} have been analyzed using this concept of charge renormalization
and in general good agreement was found for monovalent counterions
provided the colloids are far away from charged plates \cite{exp2}.

\subsection{The solvent-renormalized Yukawa model (SYM)}

This approach is a generalization of the Poisson-Boltzmann cell
modell \cite{Alexander} to the presence of a granular solvent.
Again, one considers a single polyion in a spherical cell, but uses
 the full HSSM to obtain the  counterion 
density ${\tilde \rho}_c$ at the cell boundary.  As in the PBYM, the associated
Yukawa pair potential has a solvent-renormalized inverse screening length $\kappa^*$
and a solvent-renormalized  charge $q_p^*$,
which, however,  
differ from that of the PBYM approach.

\section{Simulation method}

We consider two large spherical polyions in a cubic simulation box of length $L$
with periodic boundary conditions, hence $\rho_p=2/L^3$.
The polyions are fixed  along the body diagonal of the cubic box. While the simulation methods
for the PM are straightforward and are described elsewhere \cite{Irene,Trigger}, a
significant volume fraction of solvent particles together with large colloidal particles
implies a huge number of solvent spheres in the simulation box.

\begin{figure}
   \epsfxsize=7.8cm %6cm
   \epsfysize=7.8cm%7cm
  ~\hfill\epsfbox{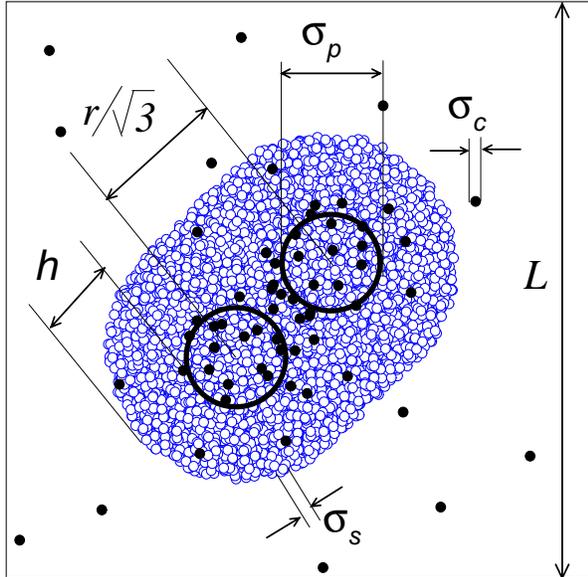}\hfill~
   \caption{View of the set-up as a downward projection of a simulation snapshot:
 Two polyions (dark open circles) in a bath of solvent particles (small hollow spheres)
contained in a spherocylindric cell  of width $h$.
The counterions  shown as small dark spheres can move in the whole
simulation box of size $L$. The distance between the
polyions is $r$, hence  the projected distance shown is $r/\sqrt{3}$.} 
     \label{fig_1}
\end{figure}

As the solvent interactions are short ranged and the solvent-averaged interactions are
only of range $\xi$, one can reduce the number of solvent particles in the simulation box
considerably using a ``solvent-bath"  method. This procedure is sketched
in Figure~\ref{fig_1} and works as follows: we define a spherocylindrical cell  around the 
colloidal pair  such that the minimal
distance $h$ from the colloidal surface to the spherocylindrical boundary is much larger
than the  hard sphere bulk correlation length $\xi$. 
The hard sphere solvent is only 
contained in this spherocylinder. As the cell volume is considerably
smaller than the volume $1/L^3$ of the whole simulation box, the number of solvent 
particles for fixed given volume fraction $\phi_s$ is drastically reduced.
No restriction is done for the counter- and salt-ions, which can move within 
the whole simulation cell. 

Care has to be taken at the artificial cell boundary.
We use the Molecular Dynamics (MD)
method calculating the particle trajectories and performing statistical  averages over 
some physical quantities. The hard sphere solvent is treated by the well-known hard sphere collision
rules.
Once a solvent particle is leaving the spherocylindric  cell  it is randomly inserted at another 
place of the cell   boundary with the same velocity. The position of
its random insertion 
has a randomly chosen distance from the spherocylindrical boundary up
to $2\sigma_s$ which avoids unphysical  
solvent layering there.

\hspace{-0.6cm} Since  the width of the cell $h$ is much larger than the hard sphere bulk
correlation length $\xi$,
the presence of the boundary has no influence on the inhomogeneous
density distribution of the solvent 
and the counterions near the colloidal surfaces. The counterion motion is implemented as follows:
Outside the spherocylindrical cell the counterions interact via a solvent-averaged
effective depletion potential. This is justified as the typical distance between
the counterions is much larger than the hard sphere bulk correlation length $\xi$.
Therefore, the correction to $V_{cc}(r)$ due to solvent layering is negligibly small anyway
for counterions outside the spherocylindrical cell. This
is not the case when salt is added as the attraction between co- and counterions
may lead to short distances where solvent depletion effects may dominate the interactions.
For  a  counterion approaching  the cell boundary, there is
an artificial  asymmetry between the solvent bath inside the cell and 
the ``vacuum" outside the cell which hinders a counterion to penetrate 
into the solvent bath. This unphysical effect is repaired in the simulation scheme
by switching off the counterion-solvent interaction for a counterion which
is penetrating from outside. Once the counterion is fully
surrounded by solvent molecules the interaction is turned on again. This
leads to a symmetric counterion crossing rate across the spherocylindrical cell boundary.

We have carefully tested the algorithm against  simulations of small systems where
the whole space was filled with solvent particles. We also tested against a situation 
of a   solvent slab between charged plates. Perfect agreement was found compared to
simulations where the whole space was filled with solvent particles.

\section{Results for the neutral case}

Let us first discuss the much simpler case of neutral polyions ($q_p=0$) under absence
of  counterions. The resulting system is just a pair of big hard spheres in a sea of small
solvent spheres. This  simple model for binary mixtures of hard sphere colloids has
gained considerable attention during the past ten years. The effective interaction between the
large spheres as induced by depletion of the small spheres in the zone intermediate
between the nearly touching big spheres has the following characteristic features
which were obtained by density functional theory \cite{depl1a,depl1b,depl1c}, computer 
simulation \cite{BibenBladon,depl2a,depl2b}
and experiments \cite{depl3a,depl3b,depl3c}:
it is attractive for nearly touching spheres, then it oscillates  with the bulk
correlation length of the hard sphere solvent $\xi$. The depletion interactions decays
to zero  exponentially with the surface-to-surface separation of the big spheres.
The characteristic decay length is again the bulk correlation length $\xi$.

Our motivation to investigate the neutral case is twofold: First, it is
a simple case which allows to test our simulation set-up. There are some
computer simulations of the depletion interaction in the literature but the range
of parameters examined is far from being complete. Second, for the solvent-averaged
primitive model (SPM) it is exactly the solvent depletion term (\ref{9})
which one has to add  
to the primitive interactions as given by eqn.(\ref{111}). Therefore, studies of the SPM
require a full knowledge of the neutral case.  

\begin{figure}
   \epsfxsize=7.8cm %6cm
   \epsfysize=7.8cm%7cm
  ~\hfill\epsfbox{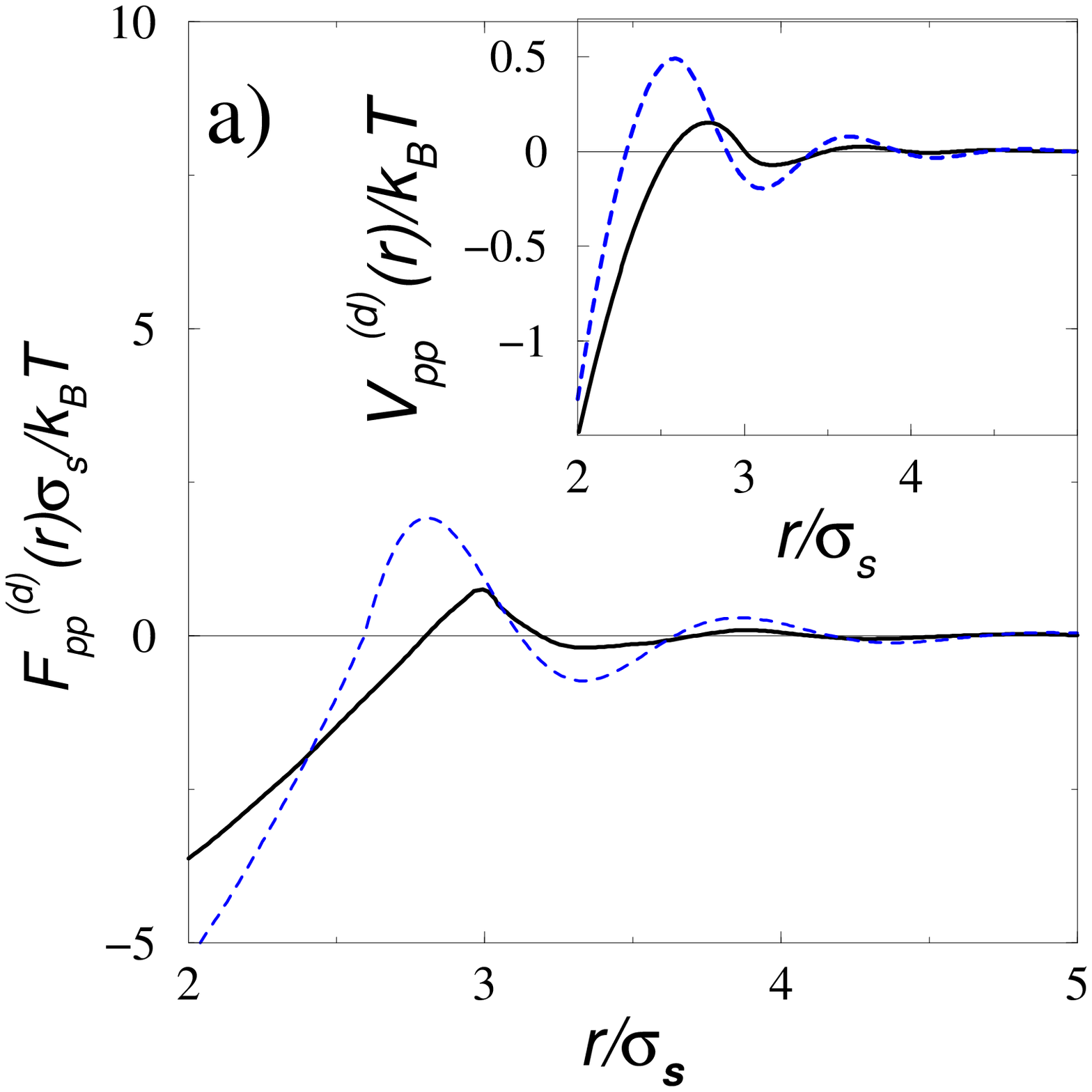}\hfill~
   \epsfxsize=7.8cm %6cm
   \epsfysize=7.8cm%7cm
  ~\hfill\epsfbox{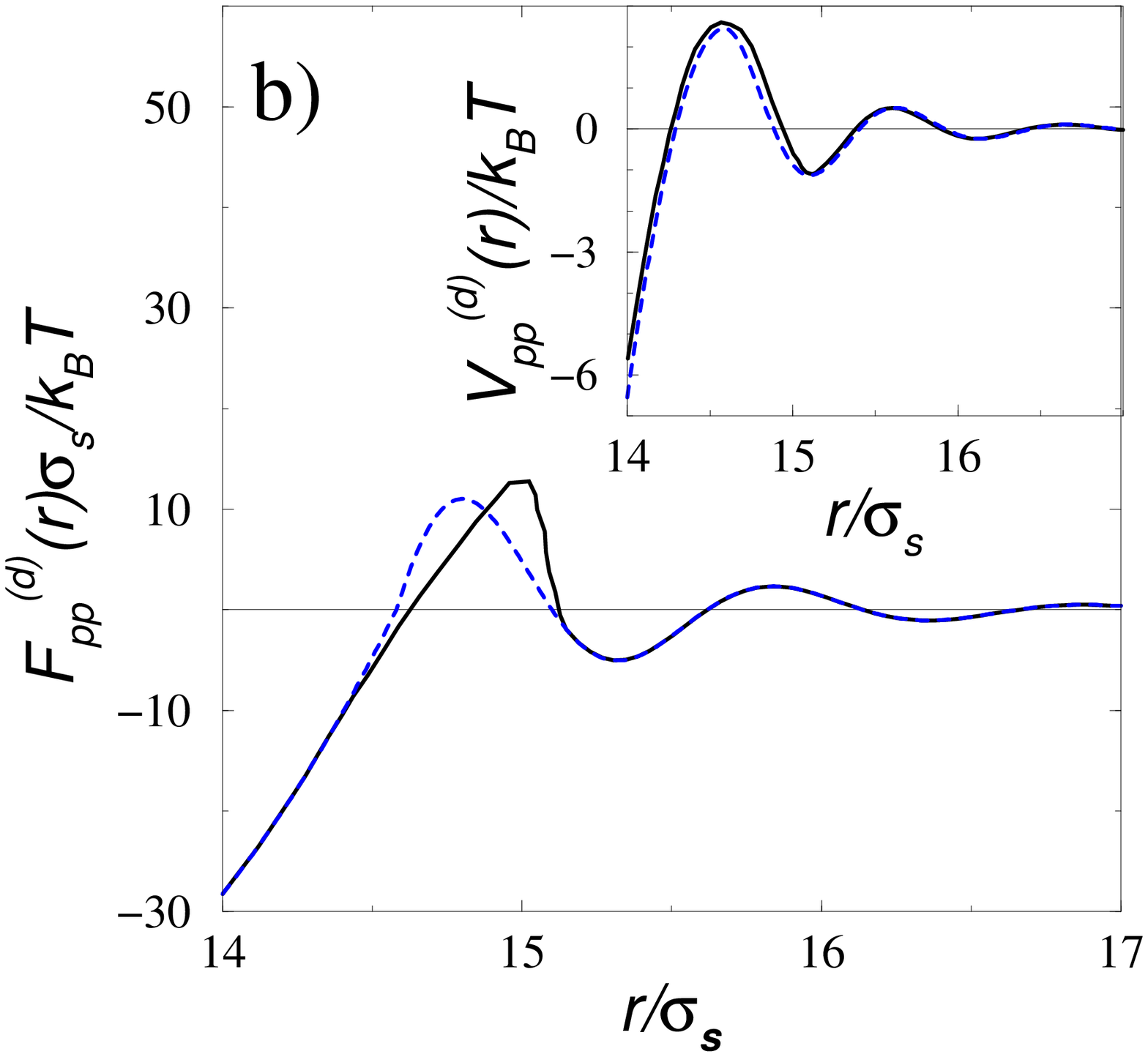}\hfill~
   \caption{Reduced distance-resolved depletion force  $F(r)\sigma_s
 / k_BT$ versus reduced 
distance $r/ \sigma_s$ between two identical neutral spheres embedded
into solvent bath of packing fraction $\phi_s=0.3$: a$)$  size
asymmetry of  $\sigma_p :  \sigma_s = 2:1$; b$)$  size
asymmetry of
$\sigma_p :  \sigma_s = 14:1$. Solid line- our
simulation results, dashed line- fitting formula of
Ref.[38]. 
%\cite{RothEvans}% .
 The inset
shows the corresponding reduced depletion potentials $ V_{pp}^{(d)}(r)/k_BT $.}
     \label{fig_2}
\end{figure}

Computer simulation results for the total effective depletion force 
${F}_{pp}^{(d)}(r)$ acting onto 
the big spheres are presented in Figure~\ref{fig_2} for two size ratios $\sigma_p/\sigma_s$
of $2$ and $14$. 
The solvent packing fraction was chosen to be $\phi_s=0.3$.
The force is projected onto the particle separation vector such that a positive sign means 
repulsion.
We note that a direct evaluation of eqn. (\ref{9}) is difficult as the solvent
density field piles up strongly at the surfaces of the big particles. A much more
effective way is to measure the momentum transfer on the fixed big spheres due to colliding
small spheres
during the Molecular Dynamics simulation.
The results are compared with a recently proposed fitting formula of Roth and Evans based
on density functional theory \cite{RothEvans}. One sees that the contact
 value of the force and the oscillations are
well-described by the theory but there are 
deviations around the first maximum. This is less apparent if the effective potential
is compared as shown in the inset of Figure~\ref{fig_2}. Obviously, the reason for that
is that the force is a derivative which is more sensitive to approximations.
In the computer  
\twocolumn[\hsize\textwidth\columnwidth\hsize\csname
@twocolumnfalse\endcsname
] 
\hspace{-0.5cm} simulation, the effective potential $V_{pp}^{(d)}(r)$  can be accessed by
integrating the distance-resolved computer simulation results of the force
\begin{equation}
V_{pp}^{(d)}(r) = - \int_{-\infty}^r dr' F_{pp}^{(d)}(r')
\label{3}
\end{equation}
%\twocolumn[\hsize\textwidth\columnwidth\hsize\csname
%@twocolumnfalse\endcsname

Also the whole set of depletion pair  potentials $V^{(d)}_{ij}(r)$  ($(ij)=(pp),(pc),(cc)$)
which are the input of a typical  SPM
simulation are presented in Figure~\ref{fig_3}. The counterion-counterion interaction is
dominated by the Coulomb repulsion also shown as a dashed line in
Figure 3c. The bare 
Coulomb repulsion between the polyions is much larger than the polyion-polyion depletion 
potential and is not shown in Figure 3a. Finally, the polyion-counterion depletion
interaction exhibits a deep
attraction near contact of the order of $k_BT$ which is of similar order than
the Coulomb attraction also shown as a dashed line on Figure 3b. This
 will have important consequences of counterion
adsorption on the colloidal surface. This effect is induced by the granularity of the solvent
and is absent in the PM.

\section{Results for the salt-free case}

\subsection{Nano-sized colloids}

Although the amount of solvent which has to be simulated explicitly has been reduced
drastically by the solvent-bath scheme, only colloidal sizes which are in the nano-domain can be
addressed on present-day computers. We have performed extensive computer 
simulation in this domain to check
carefully the different approaches. We find that the SPM describes the full simulation
data of the HSSM very well. Larger colloidal sizes are thus  only accessible within 
the SPM and discussed in chapter V.B.

In our simulations, we fixed   $T=298^oK$ and 
$\epsilon=81$ (water at room temperature) with $\sigma_s=3\AA$, $\phi_s=0.3$
such that $\xi$ is about $3\sigma_s$. We varied the polyion charge and size and
the counterion diameter $\sigma_c$. The width of the spherocylindrical cell $h$ is $10\sigma_c$ 
such that typically
$N_s=25.000-30.000$  solvent hard spheres are simulated. 

We have basically calculate two quantities: first, as a reference, we have
calculated the spherically averaged  counterion density profile $\rho_c( r)$ 
around a single polyion where $r$ is the distance from the polyion center.
The simulation was done in a cubic box of reduced length $L/2^{1/3}$ with periodic boundary conditions 
in order to reproduce the colloidal packing fraction $\phi_p$. Second, our target quantity is
the total force $F(r)$ acting onto a polyion for a given colloid-colloid separation  $r$. 
This effective  force $F(r)$ is the sum of four different contributions:\\
 i) the direct Coulomb repulsion
as embodied in $V_{pp}(r)$ (note that  all the periodic images contribute
to the total force),\\
 ii) the counterion screening resulting from the averaged
Coulomb force of counterions acting onto the polyions,\\ iii)
the counterion depletion term arising from the 
hard sphere part of $V_{pc}(r)$,\\
 iv) the solvent depletion force.

Explicit results for  $F(r)$  are presented
in Figure~\ref{fig_4} where the solvent and the counterion diameter were chosen to be equal
and the counterion were monovalent.
%\twocolumn[\hsize\textwidth\columnwidth\hsize\csname
%@twocolumnfalse\endcsname
\begin{figure}
   \epsfxsize=8cm %6cm
   \epsfysize=7.cm%7cm
  ~\hfill\epsfbox{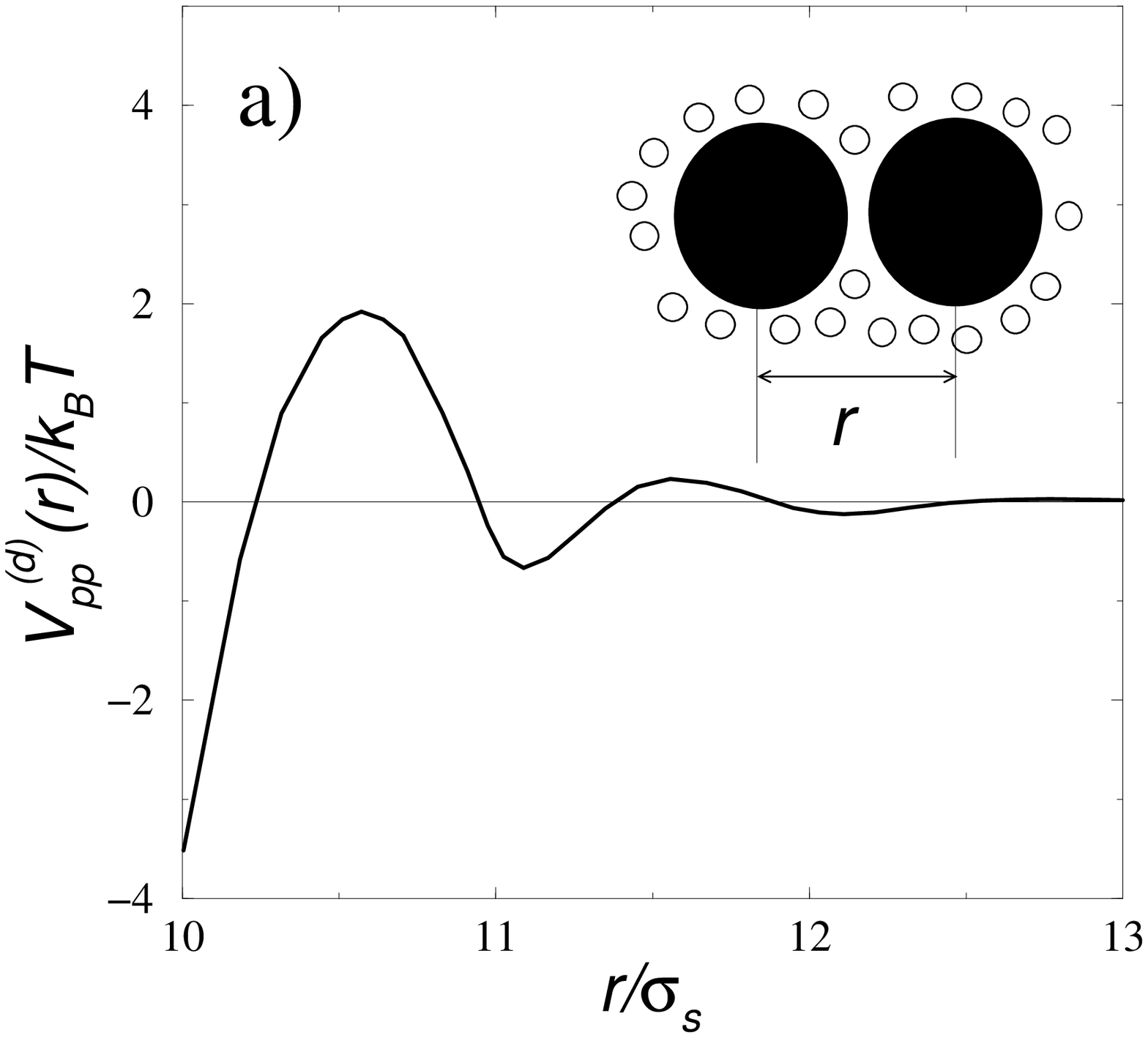}\hfill~
\end{figure}
\begin{figure}
   \epsfxsize=8cm %6cm
   \epsfysize=7.cm%7cm
  ~\hfill\epsfbox{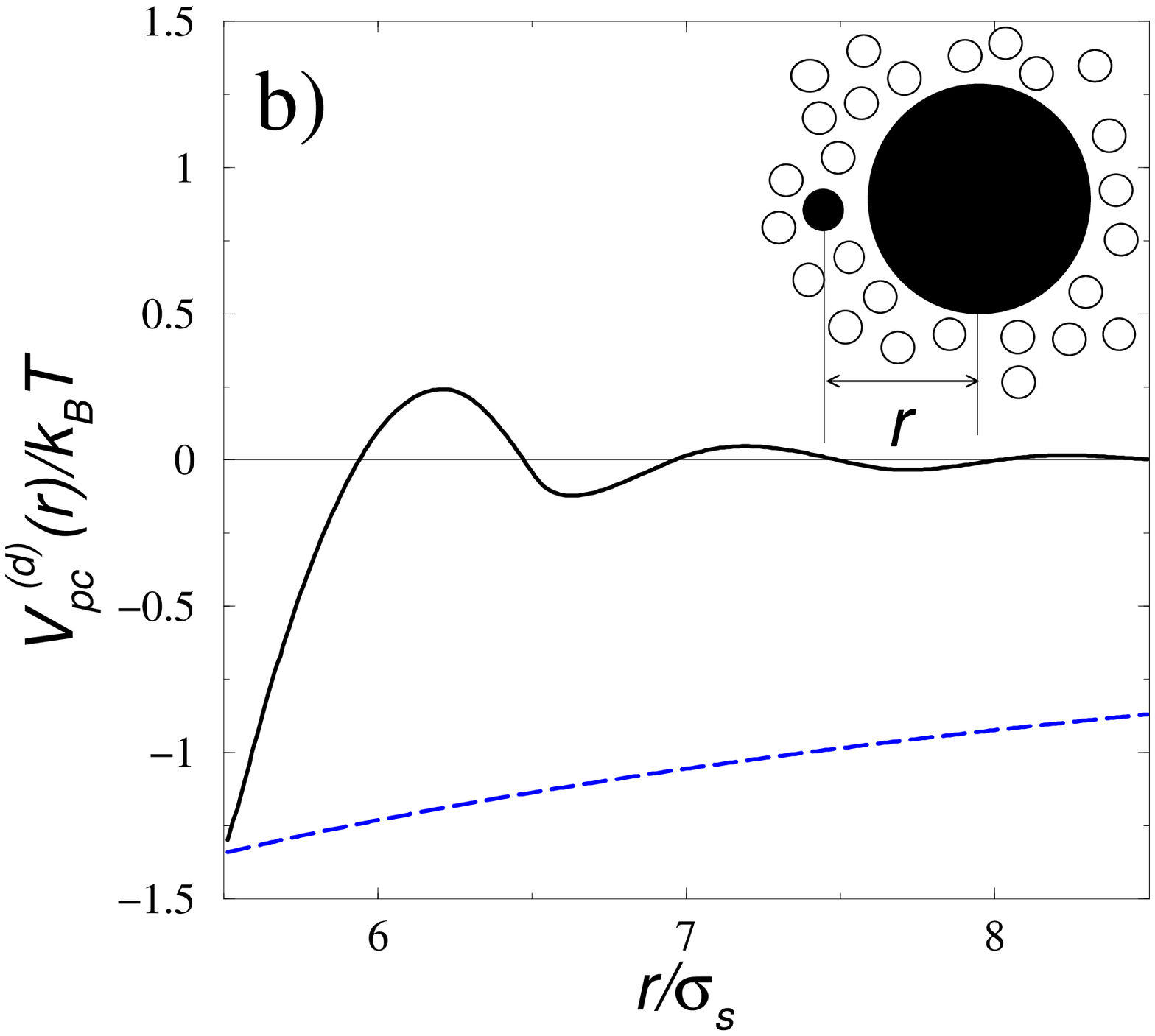}\hfill~
\end{figure}
%] 
\begin{figure}
   \epsfxsize=8cm %6cm
   \epsfysize=7.cm%7cm
  ~\hfill\epsfbox{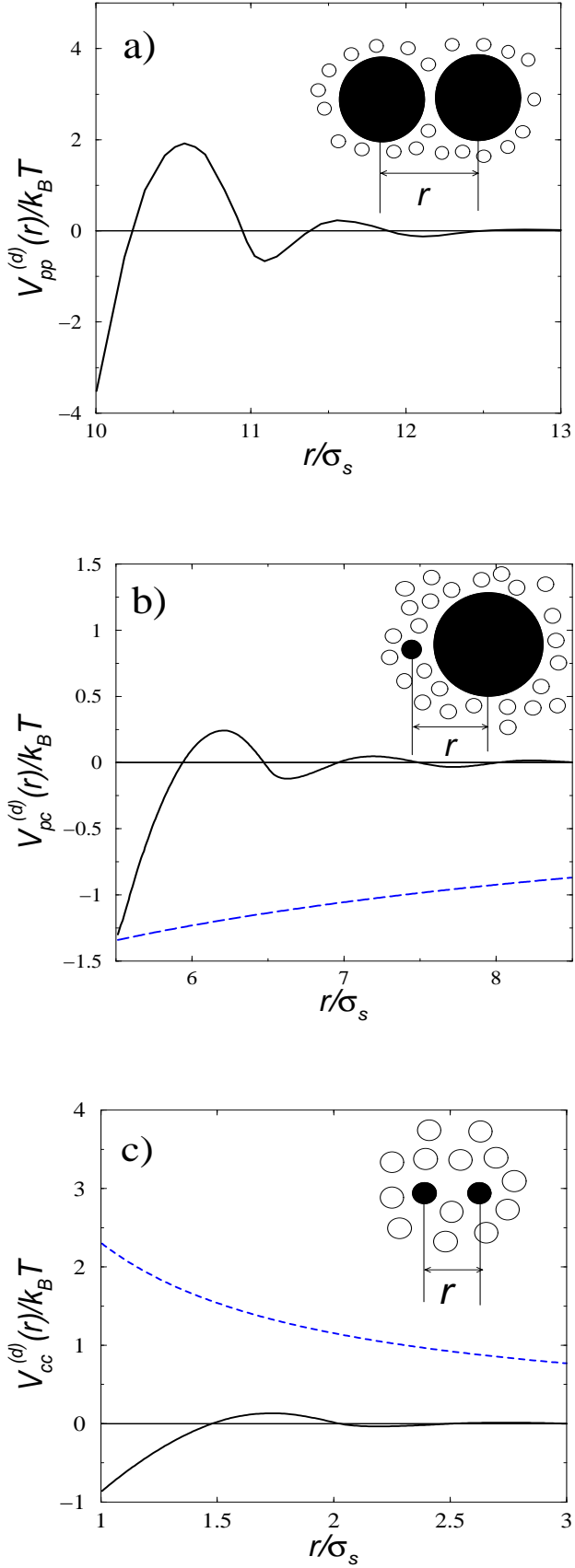}\hfill~
   \caption{Reduced  depletion potentials $V_{ij}^{(d)}(r)
/k_BT$ ($(ij)=(pp),(pc),(cc)$) versus reduced distance $r/\sigma_s$:
a) polyion-polyion depletion with $\sigma_p/\sigma_s=10$;
b) polyion-counterion depletion with $\sigma_p:\sigma_c:\sigma_s=10:1:1$
a) counterion-counterion depletion with $\sigma_c/\sigma_s=1$
The  solvent  packing fraction is $\phi_s=0.3$. The inset shows
the situation. 
The dark spheres correspond to the pair of charged
particles, the solvent are the hollow spheres. The $x$-axis starts
for touching particles. For comparison we have also included the Coulomb interaction of the PM
as dashed lines in b) and c) for $q_p=-32e$,$q_c=1e$ and $\epsilon=81$. 
Note that, the polyion-counterion Coulomb
     potential in plot 3b is reduced by a factor of 1/10.}
     \label{fig_3}
\end{figure}
%] 

\begin{figure}
   \epsfxsize=8cm %6cm
   \epsfysize=8.5cm%7cm
  ~\hfill\epsfbox{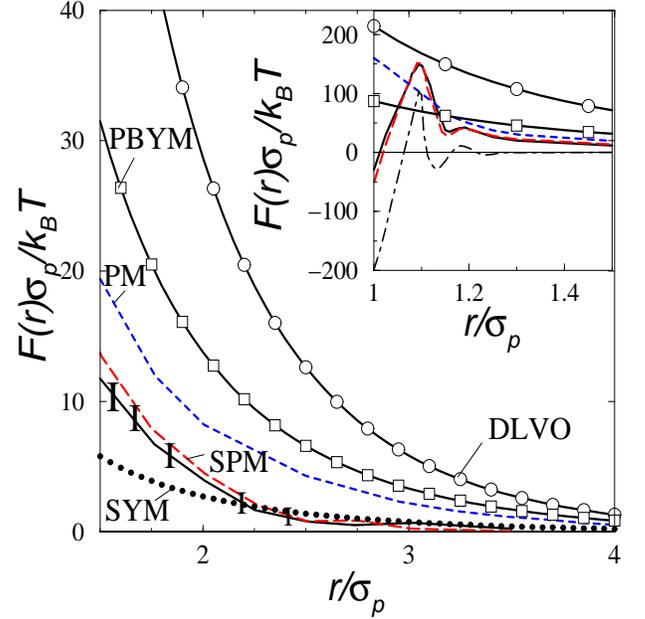}\hfill~
 \caption{Reduced distance-resolved force $F(r)\sigma_p /k_BT$ versus reduced
distance $r/\sigma_p$. The inset shows the same for nearly touching
polyions of molecular distances.
 The simulation parameters are: $q_c=1e$, $q_p=-32e$, $\epsilon=81$,
  $\sigma_p : \sigma_c : \sigma_s
= 10:1:1$, $\phi_p=5.8\times 10^{-3}$. Solid line with error bars:
  -HSSM; long-dashed line: SPM; short-dashed
 line: PM; open circles: DLVO theory; open squares: PBYM theory;
 dotted line : SYM theory; dot-dashed line in inset: solvent depletion
force (for comparison).}
    \label{fig_4}
\end{figure}

\begin{figure}
   \epsfxsize=7.8cm %6cm
   \epsfysize=7.8cm%7cm
  ~\hfill\epsfbox{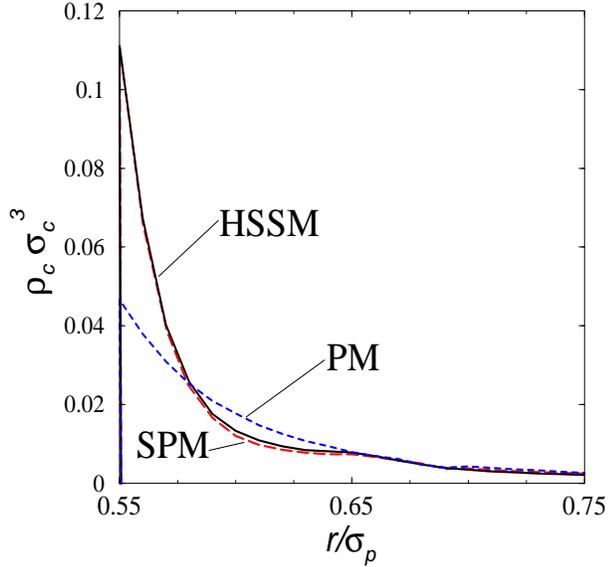}\hfill~
   \caption{Reduced counterion density profile $\rho_c\sigma_c^3$ around a single
polyion versus reduced distance $r/\sigma_p$ from the polyion center.
The parameters and the line types are as in  Figure ~\ref{fig_4}.}
    \label{fig_5}
\end{figure}

 The force exhibits oscillations for molecular distances due to
solvent and counterion layering and is repulsive for
larger distances. The SPM yields surprising agreement with the HSSM
describing even the molecular oscillations for nearly touching polyions, see the 
inset of Figure~\ref{fig_4}, while the PM overestimates the force considerably. This can be attributed
to the fact that the SPM incorporates the additional counterion accumulation
at the colloidal surface due to the hydration or solvent depletion. This can clearly
be seen in the counterionic density profile around a polyion as shown
in Figure~\ref{fig_5} which piles up near the colloidal surface.
While this accumulation is quantitatively described by the SPM it is absent in the 
 ordinary PM. The PBYM and DLVO theory lead to forces which strongly overestimate the 
HSSM data.

We have further tested the frequently invoked ``{\it superposition principle}" which approximates
the total force as a sum of the PM and the depletion term. Its comparison to
the full HSSM data is given in   Figure 6a. The first maximum of the total force is semi-quantitatively 
reproduced but the superposition principle predicts a second maximum which is too sharp
as compared to the HSSM data. This becomes even worse for a doubled counterion diameter
of $6\AA$ where the superposition predicts a secondary maximum which is completely
absent in the HSSM data, see Figure 6b. The physical reason for that is that 
the counterion layering coupled to the solvent degrees of freedom becomes relevant for these
distances.

The forces for a  doubled counterion diameter $\sigma_c$
are presented in Figure~\ref{fig_7}. For small distances (except for touching), the
 PM yields larger forces as compared to Figure 4, as the counterion repulsion is stronger
which reduces  screening. In the HSSM and SPM, on the other hand, also the polyion-counterion 
depletion attraction is getting stronger, such that the total  polyion screening is
practically unaffected. Of course, the PBYM and DLVO theory yield results which are
insensitive to the counterion diameter.

Furthermore we have investigated the case of stronger Coulomb coupling
by considering divalent counterions.
Explicit data are shown in Figure~\ref{fig_8}. There is overscreening of polyions resulting
in a mutual attraction between like-charged polyions. We emphasize
that it is the electrostatic term of the counterions that produces
the attraction but not the counterion or solvent depletion term. 
%\twocolumn[\hsize\textwidth\columnwidth\hsize\csname
%@twocolumnfalse\endcsname
\begin{figure}
   \epsfxsize=7.8cm %6cm
   \epsfysize=7.8cm%7cm
  ~\hfill\epsfbox{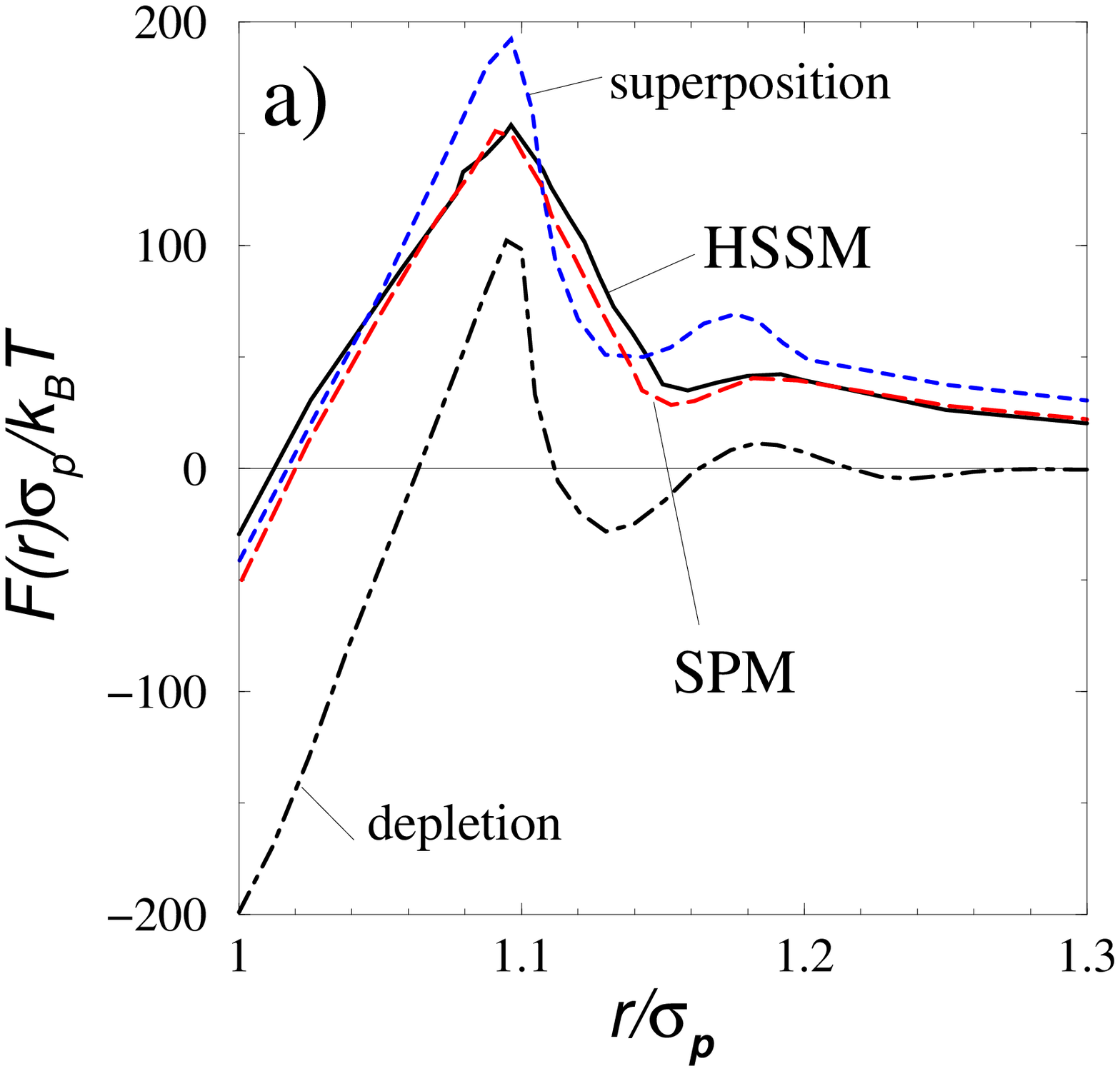}\hfill~
\end{figure}
\begin{figure}
   \epsfxsize=7.8cm %6cm
   \epsfysize=7.8cm%7cm
  ~\hfill\epsfbox{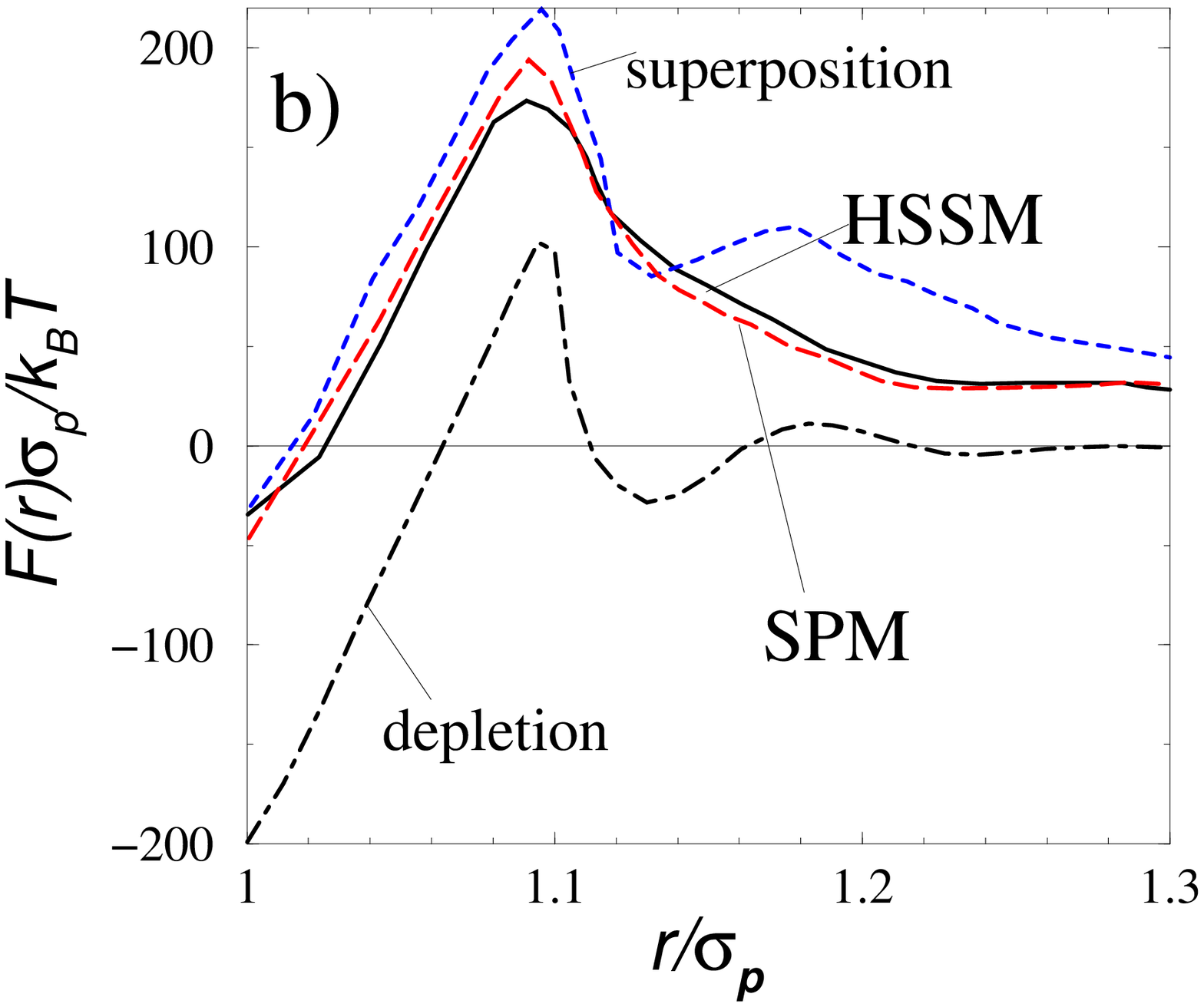}\hfill~
 \caption{Test of the superposition principle:
reduced distance-resolved force $F(r)\sigma_p /k_BT$ versus reduced
distance $r/\sigma_p$ for two different counterion sizes: a$)$
 $\sigma_c=3\AA$, b$)$ $\sigma_c=6\AA$. The force predicted by the
superposition principle is the short-dashed
 line. The other parameters
 and notations are the same as in Figure~\ref{fig_4}.}
    \label{fig_6}
\end{figure}
%] 
\hspace{-0.6cm} Nearly
every counterion is in the presence of the colloidal surfaces, as demonstrated
by the counterionic density profile shown in Figure~\ref{fig_9} where the piling-up
of counterions near the colloidal surface is much stronger. The
attractive force   has a range of
several polyion diameters. Again the SPM perfectly reproduces the
forces. The PM 
(and also the PBYM and DLVO theory), on
the other hand, yield  repulsion. This demonstrates that a discrete solvent
has a profound influence on  the effective interactions.
\begin{figure}
   \epsfxsize=8cm %6cm
   \epsfysize=9cm%7cm
  ~\hfill\epsfbox{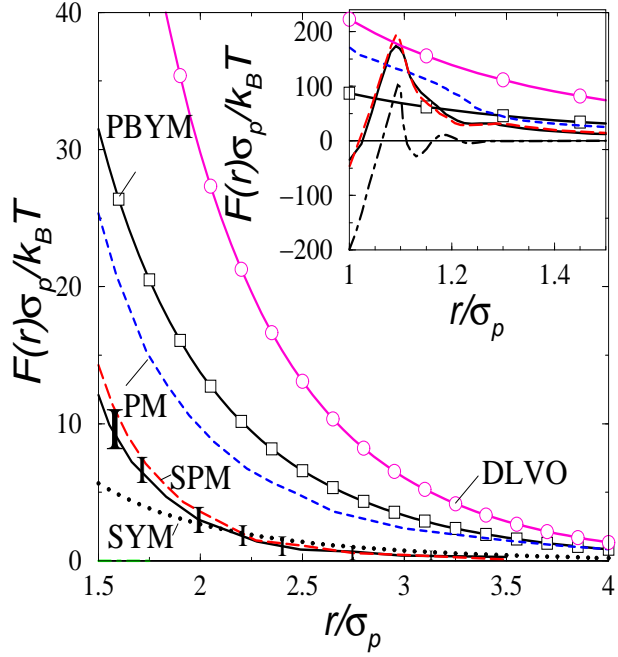}\hfill~
 \caption{Same as in Figure~\ref{fig_4} but now for a double counterion diameter
  such $\sigma_p :\sigma_c :\sigma_s = 10:2:1$.}
    \label{fig_7}
\end{figure}

\begin{figure}
   \epsfxsize=8cm %6cm
   \epsfysize=9cm%7cm
  ~\hfill\epsfbox{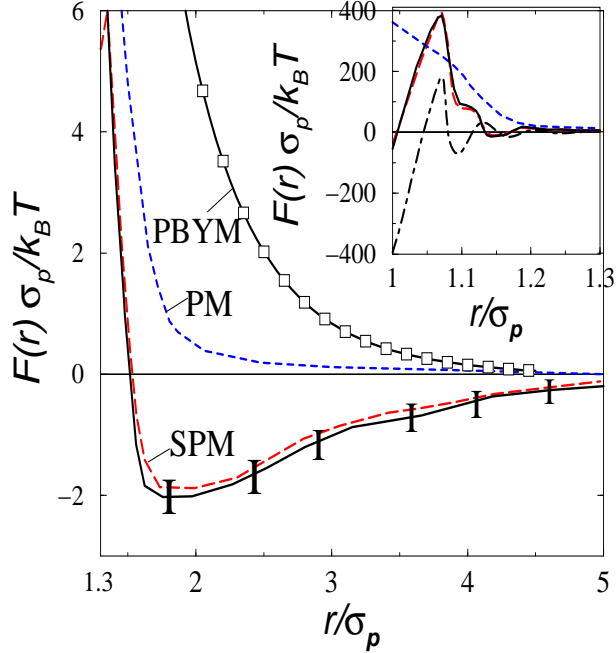}\hfill~
 \caption{Same as in Figure~\ref{fig_7} but now for
divalent counterions and  
$\sigma_p : \sigma_c : \sigma_s = 14:2:1$. The further parameters are  
$\vert q_p/q_c \vert =32$ and  $\phi_p=5.8\times 10^{-3}$.} 
    \label{fig_8}
\end{figure}
We finally discuss the validity of the  solvent-renormalized Yukawa
model (SYM). 
Computer simulations have been performed for a single polyion in a
spherical cell 
and the counterion boundary density was calculated. The boundary of the cell
was not hard but counterions leaving the cell were inserted at the opposite
side of the cell. Again a smaller spherical solvent bath around the polyions with a width $h$
was used, see Figure~\ref{fig_10} for the set-up and a projected simulation snapshot. 
As Figures 4 and 7 show,
 the SYM  is indeed a reasonable description of the forces for large
 distances.  
We also remark that the SPM and the HSSM yield the same counterion density at the
boundary of the spherical cell needed as an input for the SYM which justifies
the usage of the SPM to get the solvent-renormalized Yukawa parameters of the SYM.

\begin{figure}
   \epsfxsize=8cm %6cm
   \epsfysize=9cm%7cm
  ~\hfill\epsfbox{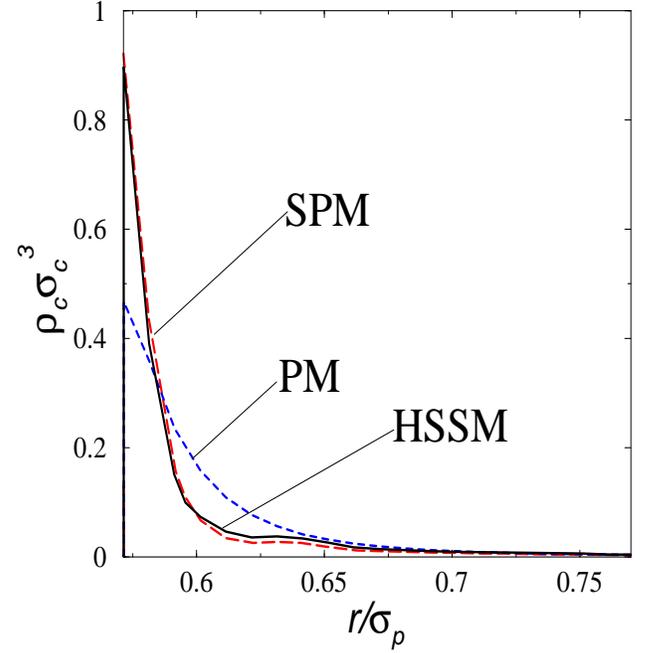}\hfill~
   \caption{Same as Figure 5 but now with the parameters of Figure~\ref{fig_8}.}
    \label{fig_9}
\end{figure}

\hspace{-0.6cm} The validity of the SYM only holds
for the case of monovalent counterions where the remaining ``free" counterions are 
responsible for the screened, repulsive force. For divalent counterions, no
free counterions are left, and a linearized screening theory  breaks down such that
an attraction cannot be encaptured by the SYM.

\subsection{Mesoscopically-sized colloids}

Our results for mesoscopically-sized colloids are based on SPM simulations
as justified in the previous chapter.
Distance-resolved colloidal forces $F(r)$ for monovalent counterions, 
a size asymmetry of
$\sigma_p : \sigma_c : \sigma_s = 370:1:1$ or $370:2:1$, and a charge ratio of $q_p/q_c=280$
are presented in Figure~\ref{fig_11ab}. These forces
 are repulsive but  much smaller than that from PM simulations.
Again, this is due to counterion accumulation near the colloidal surface as induced by the 
additional solvent depletion attraction. As 
 the corresponding potential energy gain is only few $k_BT$, this depletion attraction is 
different from chemisorption of counterions. 
The solvent-renormalized Yukawa model (SYM) leads to forces which are very similar to the SPM 
over the whole range of distances explored while the PM overestimates
the forces. 
\begin{figure}
   \epsfxsize=8cm %6cm
   \epsfysize=8cm%7cm
  ~\hfill\epsfbox{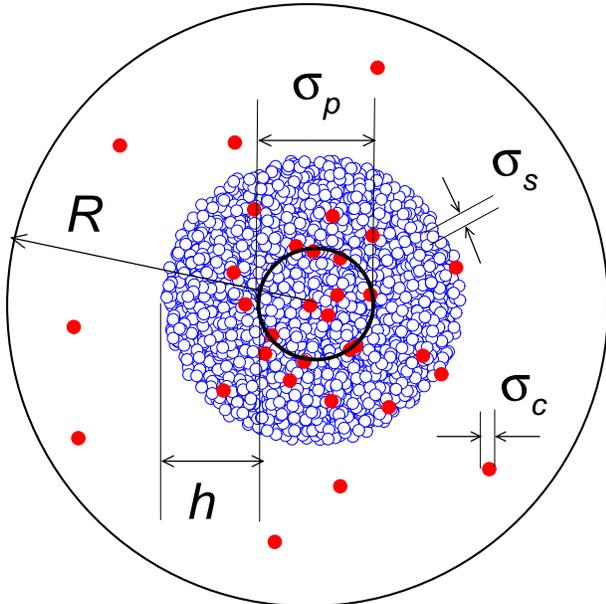}\hfill~
   \caption{View of the set-up and projected simulation snapshot for a single polyion  in a
spherical Wigner-Seitz-cell. The  polyion is shown 
as dark open circle in the cell  in a bath of solvent particles
 (small hollow spheres) contained in a spherical  cell  of width $h$.
The counterions are shown as small dark spheres. $R$ is
 cell radius.}
     \label{fig_10}
\end{figure}

The traditional meaning of the ``bare" charge $q_p$ in the PM is not the full polyion charge
but a smaller  charge which results from a polyion charge reduction by strongly
adsorbed (or condensed) counterions. This picture can also be tested against  our results. We first
have calculated the average number of counterions in a molecular shell around the colloids
of width $\xi$. If the polyion charge is reduced by this amount and the PM is used
to predict the effective interaction, the resulting force still
overestimates the HSSM data, 
 see the open diamonds in Figure 11a. In order to fit these data satisfactorily, one has 
to assume an unphysically large width of $5\xi$ to get a charge reduction that reproduces the SPM data.
Hence the PM cannot be justified even with a polyion charge reduction. The reason for that
are the weak hydration forces which are quite different from chemisorption providing
a strong counterion binding with an energy gain of hundreds or thousands of $k_BT$.
Furthermore, an arbitrary splitting into a fraction of condensed counterions and ``free"
counterions described by DLVO, Poisson Boltzmann or any other local density functional 
theory is not possible: near the
colloidal surface the electric double layer is highly correlated such that fixing
a fraction of counterions gives a completely different picture. Only if the fraction
of free counterions is determined within an approach that includes
all these correlations (as in the SYM), a linearized screening theory far away
from the colloidal surfaces is justified.

%\twocolumn[\hsize\textwidth\columnwidth\hsize\csname
%@twocolumnfalse\endcsname
\begin{figure}
   \epsfxsize=7.8cm %6cm
   \epsfysize=7.8cm%7cm
  ~\hfill\epsfbox{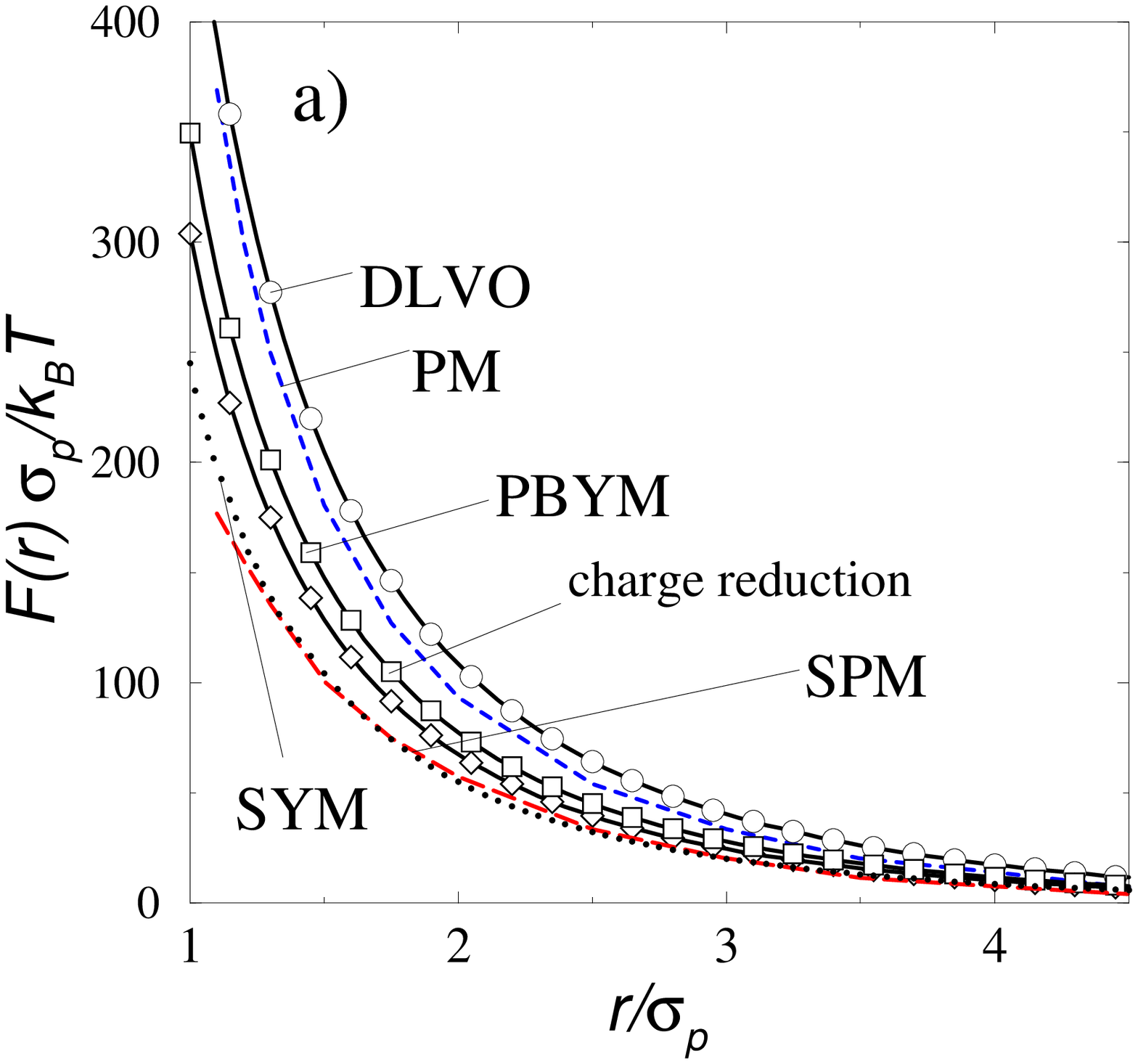}\hfill~
\end{figure}
\begin{figure}
   \epsfxsize=7.8cm %6cm
   \epsfysize=7.8cm%7cm
  ~\hfill\epsfbox{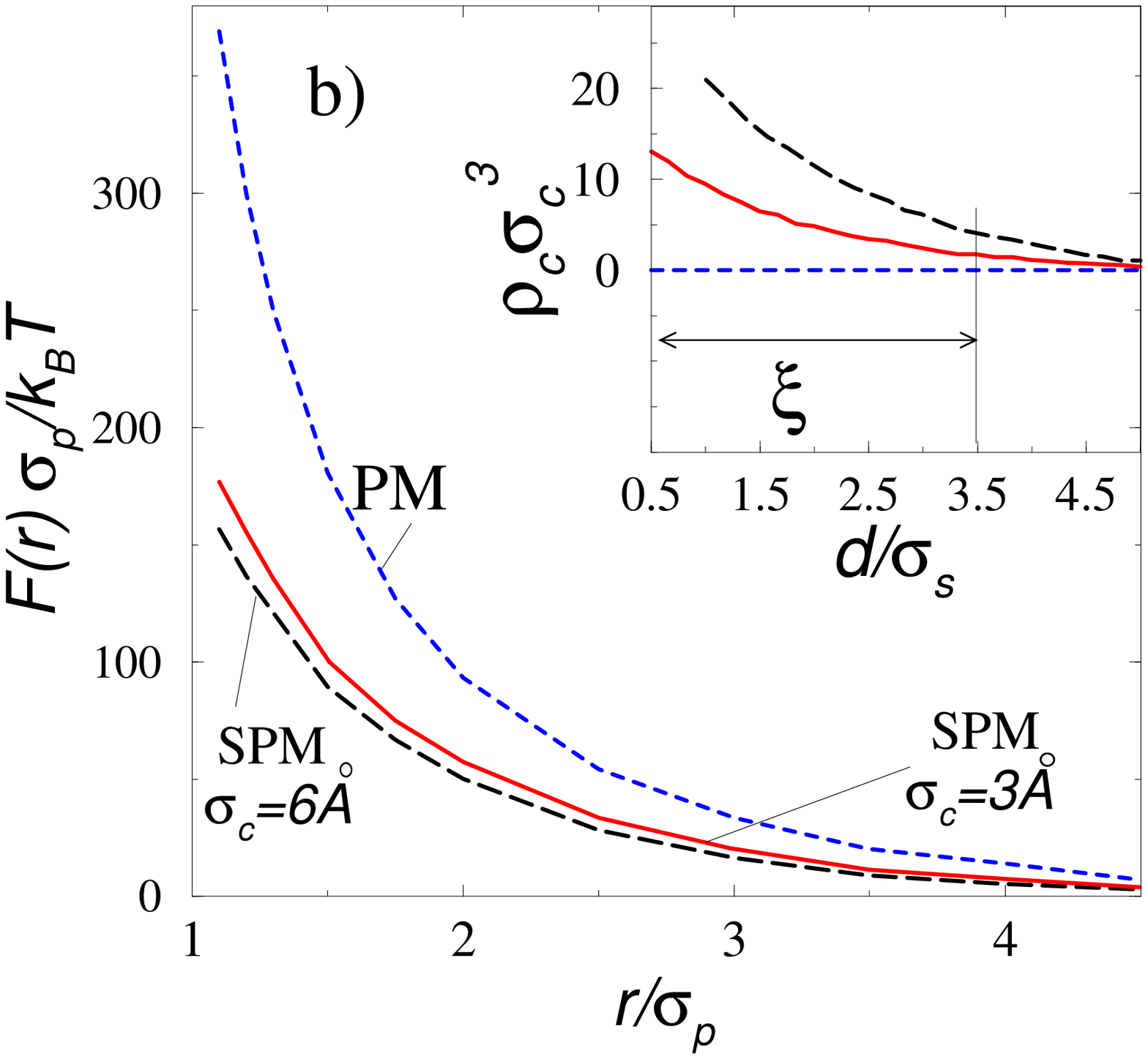}\hfill~
   \caption{Reduced distance-resolved force $F(r)\sigma_p /k_BT$ versus reduced
distance $r/\sigma_p$ for larger polyions, 
$\sigma_p:\sigma_c:\sigma_s = 370:1:1$, $\vert q_p/q_c \vert =280$,
$\phi_p=2.3 \times 10^{-3}$, and 
monovalent counterions. 
a$)$   Long-dashed line: SPM; short-dashed
 line: PM; open circles: DLVO theory; open squares: PBYM;
open diamonds: PM with charge reduction, 
 dotted line : SYM.
b$)$  Long-dashed line: SPM for a doubled counterion diameter $\sigma_c=6\AA$; solid line: 
 SPM for $\sigma_c=3\AA$; dashed line: PM. The inset  shows the corresponding
reduced counterion density
profile in the vicinity of a single polyion versus reduced distance
$d/\sigma_s$, d being the distance from polyion surface.} 
    \label{fig_11ab}
\end{figure}
%] 
This consideration casts some doubts to  some recent theories where such a partitioning
of counterions is an input, see e.g. Refs.\  \cite{Levine,Schlosski}.
Further SPM results for a doubled counterion diameter are presented in Figure 11b.
As the counterion depletion force is getting stronger for a large counterion, the force is
getting smaller, compare the full and dashed line in Figure 11b. The PM 
(short-dashed line in Figure 11b), on the other hand,
is practically insensitive to a change of the counterion diameter except very close 
to the colloidal surfaces. This picture gains further support from the counterionic 
density profiles around  a single polyion shown in the inset of Figure 11b for distances
very close to the colloidal surface. A layer of condensed but still mobile counterions
close to the surfaces is present in the SPM which is absent in the PM. The larger
the counterion diameter the more counterions are in this layers as the depletion is getting stronger.

\begin{figure}
   \epsfxsize=7.8cm %6cm
   \epsfysize=7.8cm%7cm
  ~\hfill\epsfbox{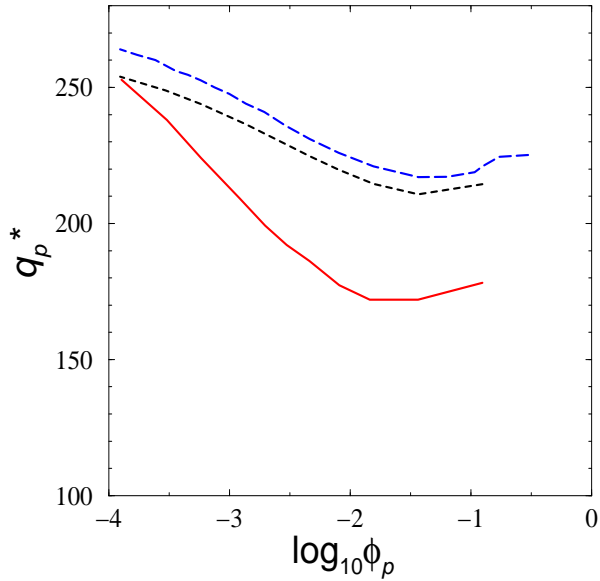}\hfill~
   \caption{Renormalized charge $q_p^*$ versus decadic 
logarithm $\log_{10}{\phi_p}$ of the polyion fraction
as obtained within in spherical cell containing a single polyion. The parameters are the same
as in Figure~\ref{fig_11ab}. Solid line: SPM; long dashed line: PBYM;
dashed line: PBYM for a fixed bare charge of $q_p =269e$.} 
    \label{fig_12}
\end{figure}

We finally discuss the solvent-renormalized charge $q_p^*$ as a function of the colloid
volume fraction  $\phi_c$ for fixed bare charge $q_p$ and compare with the prediction of the traditional
charge renormalization approach within in Poisson Boltzmann cell theory (PBYM) \cite{Alexander}.
Simulation data for $q_p^*$ 
based on the SPM in a spherical cell are shown on the full line  in Figure~\ref{fig_12}. The
renormalized charge is smaller than the bare charge and
behaves non-monotonic with the particle density. The non-monotonicity is stable
with respect to added salt and is related to a non-monotonic counterion density
at the cell boundary  as a function of density. It can be understood as follows:
For extremely high packing fractions the spherical cell accessible for the counterions
is a very thin shell across  which the polyion-counterion attraction varies slowly.
Due to the rapidly decreasing volume accessible for the counterions,
 the boundary counterion density becomes larger for increasing $\phi_p$,
see the volume fraction correction in Refs.\  \cite{Benzing,Alan}.
On the other
hand, for very small $\phi_p$, entropy of counterions will force them to cover the whole
accessible space. The counterion density at the cell boundary will increase
for decreasing $\phi_p$ getting  close
to the average density in the limit  $\phi_p\to 0$.
We remark, however, that
the non-monotonicity occurs at high polyion packing fractions of order $\phi_p\approx 0.05-0.2$
where the approximation of a spherical cell becomes questionable.
 
The PBYM  for a fixed bare
charge leads to larger values (long-dashed line in Figure 12)
which still  correctly describe the trend and 
the non-monotonicity. If the SPM data for the smallest colloid concentration are
taken as a benchmark, a bare charge of $q_p=269e$ is necessary to reproduce the 
same renormalized charge within the Poisson-Boltzmann theory. This procedure
is in strong analogy with  interpreting an experiment where the charge is a fit parameter
to describe the structural data. Starting from
this bare charge and changing the colloidal density, the PBYM predicts a similar
trend for the renormalized charge (short-dashed line in Figure 12)
but the actual numbers are different. This is consistent
with experiments on strongly deionized colloidal samples which were successfully
interpreted using a Poisson-Boltzmann renormalized colloidal 
charge \cite{Haertl,Palberg1,Palberg2}.

\section{Effects of added salt}
 
\begin{figure}
   \epsfxsize=7.8cm %6cm
   \epsfysize=7.8cm%7cm
  ~\hfill\epsfbox{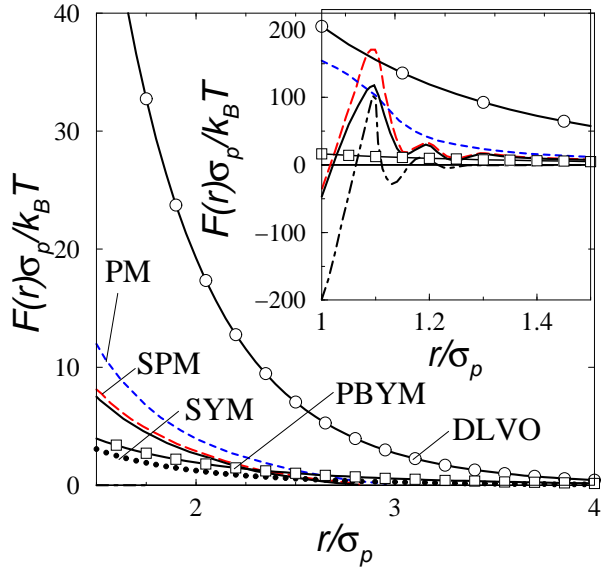}\hfill~
   \caption{Same as in Figure~\ref{fig_4} but now for added
monovalent salt, $c_s= 0.022$Mol/l.}
    \label{fig_13}
\end{figure}

Within the HSSM, the  salt ions enter as charged hard spheres. For simplicity,
we have considered a situation where the salt ions are monovalent and
have the same diameter as the counterions and the solvent. Results for the effective
interactions for a case with added salt are presented in Figure~\ref{fig_13}. As expected,
the salt ions provide an additional screening such that the forces
are less repulsive than in the salt-free case (compare with
Figure~\ref{fig_4}). 
The SPM reproduces the full HSSM data for intermediate distances but there are
deviations for molecular distances. This is in contrast to the salt-free case where
good agreement between the SPM and the HSSM was found even for small distances.
The physical reason for this is that the pair potential decomposition which is the basic
approximation of the SPM breaks down for nearly touching polyions as important
configurations are paired
microions squeezed between the polyions. This is a manifest  many-body situation 
beyond the pair level. Still for  two well-separated polyions or a single polyion,
the SPM and the HSSM yield similar results for the counterion density field or the 
colloidal forces.

\begin{figure}
   \epsfxsize=7.8cm %6cm
   \epsfysize=7.8cm%7cm
  ~\hfill\epsfbox{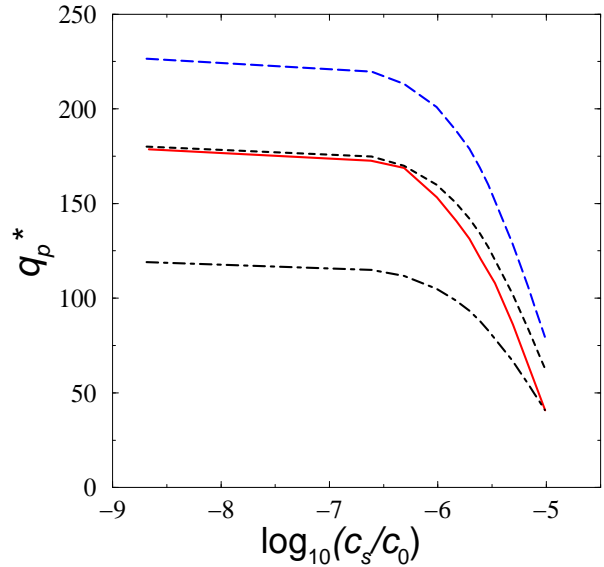}\hfill~
   \caption{Renormalized charge $q_p^*$ versus decadic 
logarithm $\log_{10}{(c_s/c_0)}$ of the salt
concentration where $c_0= 1$Mol/l is a reference salt concentration.
 The parameters are the same
as in Figure~\ref{fig_11ab} but now $\phi_p=8\times 10^{-3}$ and
$\sigma_c=3\AA$.   Solid line: SPM; long dashed line: PBYM;
dashed line: PBYM for a fixed bare charge of $q_p =210e$; 
dot-dashed line: PBYM for a fixed bare charge of $q_p =130e$.} 
    \label{fig_14}
\end{figure}

As in the salt-free case, we have  tested the PBYM. The solvent-renormalized
charge $q_p^*$ as obtained from the SPM is plotted as a solid line in Figure~\ref{fig_14} versus salt
concentration. It  decreases for increasing salt concentration. The PBYM for the same
bare charge of $q_p=280e$ yields the same trend as obtained in earlier
investigations \cite{Falk}, see the long-dashed line in Figure 14. If
scaled by  

\hspace{-0.6cm} using the SPM data
for the salt-free case 
as a benchmark (short-dashed line in Figure 14), the trend obtained in the PBYM is almost the same as that in the SPM.
This explains the success of fitting experimental data \cite{Bucci,Haertl,Palberg1,Palberg2}
by using the PBYM for real colloidal samples which typically contain a lot of added salt.
If the SPM data for a high concentration of added salt is used as benchmark point, the
PBYM predicts much smaller renormalized charge upon deionization (dot-dashed line in Figure 14).

\section{Comments on other mechanisms for polyion-polyion attraction}

\begin{figure}
   \epsfxsize=7.8cm %6cm
   \epsfysize=7.8cm%7cm
  ~\hfill\epsfbox{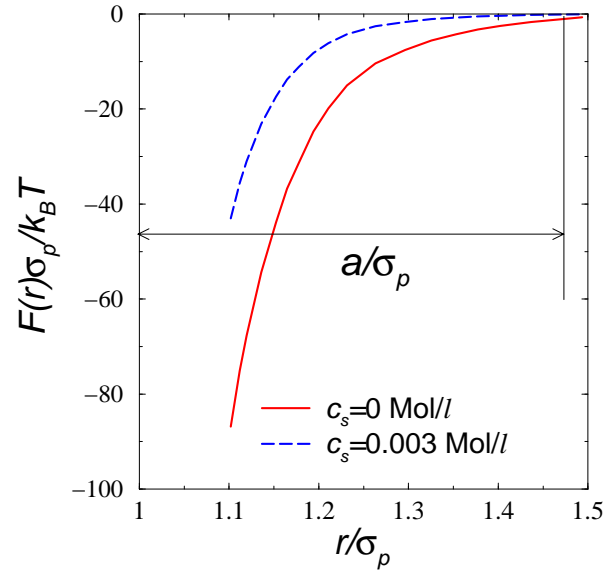}\hfill~
\caption{Reduced distance-resolved force $F(r)\sigma_p /k_BT$ versus reduced
distance $r/\sigma_p$.  The counterions are divalent and  the size
asymmetry  is 
$\sigma_p : \sigma_c : \sigma_s = 33:1:1$. The further parameters are  
$\vert q_p/q_c \vert =16$, $\epsilon=5$ and  $\phi_p=1.6\times 10^{-2}$.
Solid line: PM without salt; long-dashed line: PM with monovalent salt 
 at
concentration 
$c_s=0.003$Mol/l. The  range
 $a /  \sigma_p$=0.476 of the depletion attraction
is also indicated.} 
    \label{fig_15}
\end{figure}

We finally comment on two other physical mechanism for mutual 
attraction between like-charge colloids. The first is the counterion
depletion mechanism 
which was found within the PM in salt-free colloidal suspensions with
strong Coulomb coupling 
(as realized by a small dielectric constant) \cite{AllahyarovPRL}. We have redone
the simulation using the same parameters as in Ref.\ \cite{AllahyarovPRL} but now
with added salt. The depletion attraction is reduced but still present, see 
 Figure~\ref{fig_15}. As put forward in Ref.\ \cite{AllahyarovPRL}, 
the range of the attraction is comparable to 
\begin{equation}
a=\sqrt{q_c/q_p}\sqrt{4\pi/\sqrt{3}}\sigma_p 
\end{equation}
which is a typical counterion distance
corresponding to the spacing of a triangular lattice on the spherical colloidal surface.
This length is also included in  Figure~\ref{fig_15}.

\begin{figure}
   \epsfxsize=7.8cm %6cm
   \epsfysize=7.8cm%7cm
  ~\hfill\epsfbox{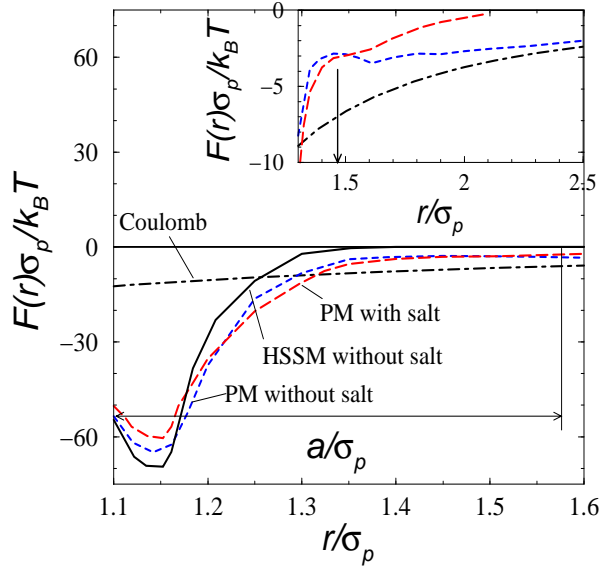}\hfill~
\caption{Reduced distance-resolved force $F(r)\sigma_p /k_BT$ versus reduced
distance $r/\sigma_p$ for divalent counterions and  
$\sigma_p : \sigma_c : \sigma_s = 10:1:1$. The further parameters are  
$\vert q_p/q_c \vert =16$, $\epsilon=20$ and  $\phi_p=5.8\times 10^{-3}$.
Solid line: HSSM without salt; long dashed line- PM result with added monovalent salt at
concentration $c_s=2.74 \times 10^{-4}$Mol/l;  
dashed line: PM without salt; dot-dashed line: pure electrostatic
interaction between a pair  of +2e and -2e ions. 
 The depletion range 
$a/ \sigma_p$=0.476 is also shown.
The long range tails of the forces are compared  in the inset. The arrow there indicates
the range $a$.}
    \label{fig_16}
\end{figure}

Unfortunately, the polyion radius used in Ref.\ \cite{AllahyarovPRL}
is too large to allow for a reasonable number of solvent particles in the solvent bath.
Therefore we  have  slightly reduced the polyion size such that the PM yields
the same counterion depletion-mediated attraction. 
Results based on the HSSM and PM are collected in Figure~\ref{fig_16}.
As can be deduced form this figure, the depletion-mediated attraction is stable 
 with respect to an explicitly added
solvent. It is further stable but reduced with respect to added salt. However,
an added solvent reduces the attraction a bit.  The physical reason
for that is that the 
solvent will prefer to stay in the counterion-free space near the colloidal surfaces
such that the solvent depletion cancels part of the counterion depletion force.

Second, we comment on the mechanism of metastable oppositely-ionized colloids leading to 
long-ranged Coulomb attraction
as found in recent salt-free PM simulations by Messina et al \cite{Messina}. We have confirmed and
reproduced this effect in our simulations for the parameters of Figure 16.
 We found, however, that the opposite ionization of
 the colloids will be immediately suppressed if salt is added. In
Figure~\ref{fig_16}, it is shown
that for large distances the salt-free PM data are close to the long-ranged Coulomb force for
 an ionization degree by one counterion (compare the dot-dashed and 
the short-dashed line in the inset of Figure 16). Once salt is added, however, the additional
microscopic ions will be attracted directly towards the ionized polyions and the long-ranged
Coulomb attraction disappears (see the long-dashed line in the inset of Figure 16). Hence,
the  mechanism of attraction due to metastable ionized states
is not stable with respect to added salt. We also remark that
metastable ionized states will disappear for separations shorter than
the characteristic depletion zone length $a$. For such close
configurations, the mutual attractive Coulomb correlations in the
counterion cloud around  both polyions will lead to a  symmetric shearing
of counterions by the two neighboring polyions. For such small separations, counterion
depletion is responsible for the attraction. 
As a function of distance, the total force is
nonmonotonic. For small distances it is dominated by counterion depletion which decays off
rapidly on the scale $a$, while
for larger distances the electrostatic resulting from the 
metastable oppositely-ionized colloids leads to a long-ranged attraction.

\section{Conclusions}

In conclusion, based on simulations of an model which contains the granularity
of the solvent explicitly, we have shown that hydration forces profoundly influence 
 the colloidal interaction. For divalent counterions, there is solvent-induced attraction 
 which is not contained in  the traditional primitive model but can be encaptured 
within a solvent-averaged primitive model (SPM).
For monovalent counterions, the forces 
 can be described by a  solvent-induced {\it charge renormalization\/}.
This picture is in agreement  with experiments on strongly deionized samples
where a Yukawa picture can be employed
provided the colloidal charge is renormalized towards a value  smaller
than the bare charge \cite{Gisler}. The trends of the renormalized charge upon
increasing the salt concentration is similar in the Poisson-Boltzmann cell model
 and the SPM which explains
why the experimental data could be well-described by using a Yukawa interaction
with a Poisson-Boltzmann-renormalized charge \cite{Haertl,Palberg1,Palberg2}. 
Still, quantitatively, there are differences between the
renormalized charges of the Poisson-Boltzmann cell model and the SPM.

Future research should focus on the role of the permanent dipole moment in a polar
solvent as modelled by dipolar hard spheres \cite{Lado,Levesque} 
or a Stockmayer liquid \cite{Dietrich0}. Also
more work has to be done to explore the role of charge regulation and chemisorption
of counterions near the colloidal surfaces. Furthermore
the dielectric discontinuity at the colloidal surfaces resulting in image
charge effects has to be explored in more detail. For all these circumstances
the concept of a renormalized polyion charge resulting in  
a  Yukawa picture should be possible provided there are free counterions left
which dominate the effective repulsion between the colloids.

We finally point out further possible applications of our simulation technique:
If used without the confining solvent-bath shell,
 our approach starts ``ab initio" and even employs the correct 
microscopic (molecular) dynamics of the solvent. Therefore,
 it could also  be used to address dynamical questions
in equilibrium and non-equilibrium. Important examples concern the motion
of poly- and counterions under the influence of an external electric field
including effects as the electrophoretic mobility \cite{mobility,m1,m2,m3}, 
ion migration \cite{Wojcik}, electro-kinetic properties \cite{Deggelmann}
 and 
electrolyte friction \cite{Schurr,electrolyte}. Our approach produces  both diffusive motion
and hydrodynamic interactions mediated by the solvent as an output. Of course, one
will not be able to simulate large time scale separations between the Brownian
and the structural relaxation time  \cite{PuseyLH}  but one should try to start
with moderate time scale separations in order to test the approximative theories
of electrophoresis.

We  thank R. Roth, M.Schmidt and T. Palberg for  helpful remarks and the DFG for financial support.


\references
\bibitem{ref1} ``Structure and Dynamics of Strongly Interacting Colloids and 
Supramolecular Aggregates in Solution" edited by S.-H. Chen, 
J. S. Huang, P. Tartaglia, NATO ASI Series, Vol. 369, Kluwer 
Academic Publishers, Dordrecht, 1992.
\bibitem{HansenLoewen} J. P. Hansen, H. L\"owen, Ann. Rev. Phys. 
Chemistry, October 2000, in press.

\bibitem{Hribar} B. Hribar, V. Vlachy, J. Chem. Phys. B {\bf 101}, 3457 (1997); 
Biophysical Journal {\bf 78}, 694 (2000).
\bibitem{Pincus} N. Gronbech-Jensen, K. M. Beardmore, P. Pincus, Physica A {\bf 261}, 74 (1998).
\bibitem{AllahyarovPRL} E. Allahyarov, I. D'Amico, H. L\"owen, Phys. Rev. 
Lett. {\bf 81}, 1334 (1998).

\bibitem{Linse} P. Linse, V. Lobaskin, Phys. Rev. Lett. {\bf 83}, 4208 (1999); 
J. Chem. Phys. {\bf 112}, 3917 (2000).

\bibitem{Messina} R. Messina, C. Holm, K. Kremer, Phys. Rev. Lett. {\bf 85}, 872 (2000).

\bibitem{LHM} H.\ L{\"o}wen, J.\ P.\ Hansen, P.\ A.\ Madden,   J.\ Chem.\ Phys. {\bf 98}, 3275
(1993). 

\bibitem{Verwey} B. V. Derjaguin, L. D. Landau, Acta Physicochim. USSR 
{\bf 14}, 633 (1941); E. J. W. Verwey and J. T. G. 
Overbeek, ``Theory of the Stability
of Lyophobic Colloids" (Elsevier, Amsterdam, 1948).

\bibitem{Alexander} S. Alexander, P. M. Chaikin, P. Grant, G. J. Morales, P. Pincus, D. Hone,
J. Chem. Phys. {\bf 80}, 5776 (1984).
\bibitem{Smith} D. E. Smith, L. X. Dang, J. Chem. Phys. {\bf 100}, 3757 (1994).
\bibitem{LIE2} M. Kinoshita, S. Iba, M. Harada, J. Chem. Phys. {\bf
105}, 2487 (1996).
\bibitem{LIE} F. Otto, G. N. Patey, Phys. Rev. E {\bf 60}, 4416 (1999); 
J. Chem. Phys. {\bf 112}, 8939 (2000).

\bibitem{planar} S. Marcelja, Colloids and Surfaces A {\bf 129-130}, 321 (1997).
\bibitem{MPB} V. Kralj-Iglic, A. Iglic, J. Physique II (France) {\bf 6}, 
477 (1996); J. Borukhov, D. Andelman, H. Orland, Phys. Rev. Lett. {\bf 79}, 
435 (1997); Y. Burak, D. Andelman, to be published, E. Trizac, J.-L. Raimbault, 
Phys. Rev. E {\bf 60}, 6530 (1999).

\bibitem{DFT} Z. Tang, L. E. Scriven, H. T. Davis, J. Chem. Phys. {\bf 100}, 4527 (1994);
L. J. D. Frink, F. van Swol, J. Chem. Phys. {\bf 105}, 2884 (1996);
T. Biben, J. P. Hansen, Y. Rosenfeld, Phys. Rev. E {\bf 57}, R3727 (1998)
C. N. Patra, J. Chem. Phys. {\bf 111}, 9832 (1999);
D. Henderson, P. Bryk, S. Sokolowski, D. T. Wasan, Phys. Rev. E {\bf
61}, 3896 (2000).

\bibitem{Henderson2} See  e.g.: D. Boda, D. Henderson, J. Chem. Phys. {\bf 112}, 8934 (2000).
\bibitem{CS} J. Rescic, V. Vlachy, L. B. Bhuiyan, C. W. Outhwaite, J. Chem. 
Phys. {\bf 107}, 3611 (1997).


\bibitem{planar2} S. Marcelja, Langmuir {\bf 16}, 6081 (2000).

\bibitem{MM} W. G. McMillan, J. E. Mayer, J. Chem. Phys. {\bf 13}, 276 (1945).

\bibitem{El1} E. Allahyarov, H. L\"owen, to be published.


\bibitem{depl10} T. Biben, J. P. Hansen, Europhys. Lett. {\bf 12}, 347 (1990).

\bibitem{depl1a} M. Dijkstra, R.  van Roij, R. Evans, Phys. Rev. Lett. {\bf 81}, 2268 (1998); 
Phys. Rev. E {\bf 59}, 5744 (1999).

\bibitem{depl1b} B. G\"otzelmann, R. Roth, S. Dietrich, M. Dijkstra, R. Evans, Europhys. Lett. {\bf 47}, 398 (1999).

\bibitem{depl1c} R. Roth, B. G\"otzelmann,  S. Dietrich, Phys. Rev. Lett. {\bf 83}, 448 (1999).

\bibitem{depl2a} R. Dickman, P. Attard, V. Simonian, J. Chem. Phys. {\bf 107}, 205 (1997).

\bibitem{depl2b} M. Dijkstra, R.  van Roij, R.  Evans,
       Phys. Rev. Lett. {\bf 82}, 117 (1999).

\bibitem{Haertl} W. H\"artl, H. Versmold, J. Chem. Phys. {\bf 88}, 7157 (1988).

\bibitem{Palberg2} F. Bitzer, T. Palberg, H. L\"owen, R. Simon, P. Leiderer, Phys. Rev. E {\bf 50}, 2821 (1994).


\bibitem{Palberg1}
T. Palberg, W. M\"onch, F. Bitzer, R. Piazza, T. Bellini, Phys. Rev. Lett. {\bf 74}, 4555 (1995).
\bibitem{exp2}
J. C. Crocker, D. G. Grier, Phys. Rev. Lett. {\bf 73}, 352 (1994); 
G. M. Kepler, S. Fraden, Phys. Rev. Lett. {\bf 73}, 356 (1994); 
D. G. Grier, Nature (London) {\bf 393}, 621 (1998).
\bibitem{Irene} I. D'Amico, H. L{\"o}wen,  Physica A {\bf 237}, 25 (1997). 
\bibitem{Trigger}  E. Allahyarov, H. L{\"o}wen, S. Trigger, Phys. Rev. E {\bf 57}, 5818 (1998).
\bibitem{BibenBladon} T. Biben, P. Bladon, D. Frenkel, J. Phys. Condensed Matter {\bf 8}, 10799 (1996).
\bibitem{depl3a} D. Rudhardt, C. Bechinger, P. Leiderer, Phys. Rev. Lett. {\bf 81}, 1330 (1998).
\bibitem{depl3b} J. C. Crocker, J. A. Matteo, A. D. Dinsmore, A. G. Yodh,
Phys. Rev. Lett. {\bf 82}, 4352 (1999).

\bibitem{depl3c} C. Bechinger, D. Rudhardt, P. Leiderer, R. Roth, S. Dietrich,
Phys. Rev. Lett. {\bf 83}, 3960 (1999).

\bibitem{RothEvans} R. Roth, R. Evans, to be published.

\bibitem{Levine} M. N. Tamashiro, Y. Levin, M. C. Barbosa, Physica A {\bf 258}, 341 (1998).
\bibitem{Schlosski}  V. I. Perel, B. I. Shklovskii, Physica A {\bf 274}, 446 (1999).
\bibitem{Benzing} W. B. Russel, D. W. Benzing, J. Colloid Interface Sciece {\bf 83}, 163 (1981).

\bibitem{Alan} A. R. Denton, H. L\"owen, Phys. Rev. Lett. {\bf 81}, 469 (1998).
\bibitem{Falk} M. J. Stevens, M. L. Falk, M. O. Robbins, J. Chem. Phys. {\bf 104}, 5209 (1996).

\bibitem{Bucci} S. Bucci, S. Fagotti, V. Dergiorgio, R. Piazza, Langmuir {\bf 7}, 824 (1991).

\bibitem{Gisler} T. Gisler, S. F. Schulz, M. Borkovec, H. Sticher, P. Schurtenberger, B. D'Aguanno, R. Klein, J. Chem. Phys. {\bf 101}, 9924 (1994).

\bibitem{mobility} M. Evers, N. Garbow, D. Hessinger, T. Palberg, Phys. Rev. E {\bf 57}, 6774 (1998).
\bibitem{m1} A. K. Gaigalas, S. Woo, J. B. Hubbard, J. Colloid Interface Science {\bf 136}, 213 (1990).

\bibitem{m2} C. S. Mangelsdorf, L. R. White, J. Chem. Soc. Faraday Trans. {\bf 88}, 3567 (1992).

\bibitem{m3} H. Ohshima, J. Colloid Interface Science {\bf 179}, 431 (1996); {\bf 188}, 481 (1997).

\bibitem{Wojcik} M. Wojcik, Chem. Phys. Lett. {\bf 260}, 287 (1996).

\bibitem{Deggelmann} M. Deggelmann, T. Palberg, M. Hagenb\"uchle, E. E. Maier, 
R. Krause, C. Graf, R. Weber, J. Colloid Interface Science {\bf 143}, 318 (1991).
\bibitem{Schurr} J. M. Schurr, Chem. Phys. {\bf 45}, 119 (1980).

\bibitem{electrolyte} G. Cruz de Leon, M. Medina-Noyola, O. Alarcon-Waess, H. Ruiz-Estrada, Chem. Phys. Letters {\bf 207}, 294 (1993); J. M. Mendez-Alcaraz, O. Alarcon-Waess, Physica A {\bf 268}, 75 (1999).

\bibitem{PuseyLH}   P. N. Pusey, in ``Liquids, Freezing and
the Glass Transition",
edited by J. P. Hansen, D. Levesque and J. Zinn-Justin (North Holland, Amsterdam,
1991).
\bibitem{Lado} F. Lado, J. Chem. Phys. {\bf 106}, 4707 (1997).
\bibitem{Levesque} J. J. Weis, Mol. Phys. {\bf 93}, 361 (1998).

\bibitem{Dietrich0} B. Groh, S. Dietrich, Phys. Rev. Lett. {\bf 72}, 2422 (1994); {\bf 74}, 2617 (1995).
\end{document}